\documentclass{naturePrep}
\usepackage{graphics,graphicx,hyperref}
\usepackage{amssymb}	

\usepackage[normalem]{ulem} 

\usepackage{caption}
\DeclareCaptionLabelSeparator{natbar}{ $\vert$ }
\def\ts     {\thinspace}   
\def\hii  {\ifmmode{{\rm H}{\rm \small II}}\else{H\ts {\scriptsize II}}\fi}
\def\ci  {\ifmmode{[{\rm C}{\rm \small I}]}\else{[C\ts {\scriptsize I}]}\fi}
\def\cii  {\ifmmode{[{\rm C}{\rm \small II}]}\else{[C\ts {\scriptsize II}]}\fi}
\def\oiii  {\ifmmode{[{\rm O}{\rm \small III}]}\else{[O\ts {\scriptsize III}]}\fi}
\def\lfir  {\ifmmode{{L}_{\rm FIR}}\else{${L}_{\rm FIR}$}\fi}
\def\lir  {\ifmmode{{L}_{\rm IR}}\else{${L}_{\rm IR}$}\fi}
\def\sigfir  {\ifmmode{{\Sigma}_{\rm FIR}}\else{${\Sigma}_{\rm FIR}$}\fi}
\newcommand{\msol}{\ensuremath{M_\odot}}
\newcommand{\lsol}{\ensuremath{L_\odot}}

\newcommand{\um}{\ensuremath{\mu\rm{m}}}
\def\kms{{km~s$^{-1}$}}              
\def\arcsec {$^{\prime\prime}$}
\def\arcmin {$^{\prime}$}

\newcommand{\hst}{\textit{HST}}
\newcommand{\hubble}{\textit{Hubble}}
\newcommand{\spitzer}{\textit{Spitzer}}
\newcommand{\mdust}{\ensuremath{M_{\rm dust}}}
\newcommand{\mgas}{\ensuremath{M_{\rm gas}}}

\newcommand{\SI}{Methods}
\newcommand{\EF}{Extended Data Fig.}

\newcommand{\sname}{SPT0311--58}
\newcommand{\muE}{1.3}
\newcommand{\muW}{2.2}
\newcommand{\muTotCont}{2.0}
\def\LirE  {\ensuremath{(4.6\pm1.2)\times10^{12}}}
\def\LirW  {\ensuremath{(33\pm7)\times10^{12}}}
\def\sfrE {540$\pm$175}
\def\sfrW {2900$\pm$1800}
\def\mstarE {\ensuremath{(3.5\pm1.5)\times10^{10}}}
\def\mgasCOw{\ensuremath{(6.6\pm1.7)\times10^{10}}}
\def\mgasCOe{\ensuremath{(1.0\pm0.3)\times10^{10}}}
\def\mgasAXELw{\ensuremath{(2.7\pm1.7)\times10^{11}}}
\def\mgasAXELe{\ensuremath{(0.4\pm0.2)\times10^{11}}}
\def\mdustAXELw{\ensuremath{(2.5\pm1.6)\times10^{9}}}
\def\mdustAXELe{\ensuremath{(0.4\pm0.2)\times10^{9}}}

\def\Arizona{1}
\def\Flatiron{2}
\def\CfA{3}
\def\Illinois{4}
\def\Diego{5}
\def\MIT{6}
\def\Marseille{7}
\def\UMKC{8}
\def\Cambridge{9}
\def\Kavli{10}
\def\KICPChicago{11}
\def\PhysicsUChicago{12}
\def\EFIChicago{13}
\def\AAUChicago{14}
\def\Dal{15}
\def\ESOGarching{16}
\def\StMary{17}
\def\Davis{18}
\def\UFlorida{19}
\def\UCL{20}
\def\Stanford{21}
\def\McM{22}
\def\UCLA{23}
\def\NRAO{24}
\def\LSST{25}
\def\MPIfR{26}
\def\IMPRS{27}
\def\Hubble{28}

\title{Galaxy growth in a massive halo in the first billion years of cosmic history}

\author{D.~P.~Marrone$^{\Arizona}$, J.~S.~Spilker$^{\Arizona}$, 
C.~C.~Hayward$^{\Flatiron,\CfA}$, J.~D.~Vieira$^{\Illinois}$, 
M.~Aravena$^{\Diego}$,
M.~L.~N.~Ashby$^{\CfA}$,
M.~B.~Bayliss$^{\MIT}$,
M.~B\'ethermin$^{\Marseille}$,
M.~Brodwin$^{\UMKC}$,
M.~S.~Bothwell$^{\Cambridge,\Kavli}$,
J.~E.~Carlstrom$^{\KICPChicago,\PhysicsUChicago,\EFIChicago,\AAUChicago}$, 
S. C. Chapman$^{\Dal}$,
Chian-Chou~Chen$^{\ESOGarching}$,
T.~M.~Crawford$^{\KICPChicago,\AAUChicago}$,
D.~J.~M.~Cunningham$^{\Dal,\StMary}$,
C.~De~Breuck$^{\ESOGarching}$,
C.~D.~Fassnacht$^{\Davis}$,
A.~H.~Gonzalez$^{\UFlorida}$, 
T.~R.~Greve$^{\UCL}$,	
Y.~D.~Hezaveh$^{\Stanford,\Hubble}$,
K.~Lacaille$^{\McM}$,
K.~C.~Litke$^{\Arizona}$,
S.~Lower$^{\Illinois}$,
J.~Ma$^{\UFlorida}$, 
M.~Malkan$^{\UCLA}$,
T.~B.~Miller$^{\Dal}$,
W.~R.~Morningstar$^{\Stanford}$,
E.~J.~Murphy$^{\NRAO}$,
D. Narayanan$^{\UFlorida}$, 
K.~A.~Phadke$^{\Illinois}$,
K.~M.~Rotermund$^{\Dal}$,
J.~Sreevani$^{\Illinois}$, 
B.~Stalder$^{\LSST}$,
A.~A.~Stark$^{\CfA}$,
M.~L.~Strandet$^{\MPIfR,\IMPRS}$,
M.~Tang$^{\Arizona}$,
\& A.~Wei\ss$^{\MPIfR}$ \\
}

\begin{document}

\maketitle

\begin{affiliations}
 \item Steward Observatory, University of Arizona, 933 North Cherry Avenue, Tucson, AZ 85721, USA 
 \item Center for Computational Astrophysics, Flatiron Institute, 162 Fifth Avenue, New York, NY 10010, USA 
 \item Harvard-Smithsonian Center for Astrophysics, 60 Garden Street, Cambridge, MA 02138, USA 
 \item Department of Astronomy, University of Illinois, 1002 West Green St., Urbana, IL 61801 
 \item N\'ucleo de Astronom\'{\i}a, Facultad de Ingenier\'{\i}a, Universidad Diego Portales, Avenida Ej\'ercito 441, Santiago, Chile 
 \item Kavli Institute for Astrophysics \& Space Research, Massachusetts Institute of Technology, 77 Massachusetts Avenue, Cambridge, MA 02139, USA 
 \item Aix Marseille Universit\'e, CNRS, LAM, Laboratoire d'Astrophysique de Marseille, Marseille, France 
 \item Department of Physics and Astronomy, University of Missouri, 5110 Rockhill Road, Kansas City, MO 64110, USA 
 \item Cavendish Laboratory, University of Cambridge, 19 J.J. Thomson Avenue, Cambridge, CB3 0HE, UK 
 \item Kavli Institute for Cosmology, University of Cambridge, Madingley Road, Cambridge CB3 0HA, UK 
 \item Kavli Institute for Cosmological Physics, University of Chicago, 5640 South Ellis Avenue, Chicago, IL 60637, USA 
 \item Department of Physics, University of Chicago, 5640 South Ellis Avenue, Chicago, IL 60637, USA 
 \item Enrico Fermi Institute, University of Chicago, 5640 South Ellis Avenue, Chicago, IL 60637, USA 
 \item Department of Astronomy and Astrophysics, University of Chicago, 5640 South Ellis Avenue, Chicago, IL 60637, USA 
 \item Dalhousie University, Halifax, Nova Scotia, Canada 
 \item European Southern Observatory, Karl Schwarzschild Stra\ss e 2, 85748 Garching, Germany 
 \item Department of Astronomy and Physics, Saint Mary's University, Halifax, Nova Scotia, Canada 
 \item Department of Physics,  University of California, One Shields Avenue, Davis, CA 95616, USA 
 \item Department of Astronomy, University of Florida, Bryant Space Sciences Center, Gainesville, FL 32611 USA 
 \item Department of Physics and Astronomy, University College London, Gower Street, London WC1E 6BT, UK 
 \item Kavli Institute for Particle Astrophysics and Cosmology, Stanford University, Stanford, CA 94305, USA 
 \item Department of Physics and Astronomy, McMaster University, Hamilton, ON L8S 4M1 Canada 
 \item Department of Physics and Astronomy, University of California, Los Angeles, CA 90095-1547, USA 
 \item National Radio Astronomy Observatory, 520 Edgemont Road, Charlottesville, VA 22903, USA 
 \item Large Synoptic Survey Telescope, 950 North Cherry Avenue, Tucson, AZ 85719, USA 
 \item Max-Planck-Institut f\"{u}r Radioastronomie, Auf dem H\"{u}gel 69 D-53121 Bonn, Germany 
 \item International Max Planck Research School (IMPRS) for Astronomy and Astrophysics, Universities of Bonn and Cologne 
 \item Hubble Fellow 
\end{affiliations}

\begin{abstract}
According to the current understanding of cosmic structure formation, the precursors of the most massive structures in the Universe began 
to form shortly after the Big Bang, in regions corresponding to the largest fluctuations in the cosmic density field\cite{springel05,cole08,behroozi13}. 
Observing these structures during their period of active growth and assembly---the first few hundred million years of the Universe---is challenging
because it requires surveys that are sensitive enough to detect the distant galaxies that act as signposts for these structures and wide
enough to capture the rarest objects. As a result, very few such objects have been detected so far\cite{riechers13,vieira13}.
Here we report observations of a far-infrared-luminous object at redshift 6.900 (less than 800 Myr after the Big Bang) that was
discovered in a wide-field survey\cite{strandet17}. 
High-resolution imaging reveals this source to be a pair of extremely massive star-forming galaxies. 
The larger of these galaxies is forming stars at a rate of 2900 
solar masses per year, contains 270 billion solar masses of gas and 2.5 billion solar masses of dust, and 
is more massive than any other known object at a redshift of more than 6. Its rapid star formation is probably triggered 
by its companion galaxy at a projected separation of just 8~kiloparsecs. 
This merging companion hosts 35 billion 
solar masses of stars and has a star-formation rate of 540 solar masses per year, 
but has an order of magnitude less gas and dust than its neighbor and physical conditions akin to those observed in lower-metallicity 
galaxies in the nearby Universe\cite{cormier15}. These objects suggest the presence of a dark-matter halo with a mass of more than
400 billion solar masses, making it among the rarest dark-matter haloes that should exist in the Universe at this epoch.
\end{abstract}

\sname\ (SPT-S~J031132-5823.4) was originally identified in the 2500 deg$^2$ South Pole Telescope (SPT) survey\cite{carlstrom11,mocanu13} 
as a luminous source (flux densities of 7.5 and 19.0\,mJy at wavelengths of 2.0~mm and 1.4~mm) with a steeply rising spectrum, indicative 
of thermal dust emission. 
Observations with the Atacama Large Millimeter/submillimeter Array (ALMA) provide the redshift of the source. The $J=6-5$ and 
$7-6$ rotational transitions of the carbon monoxide molecule, along with the $^3P_2-^3P_1$ fine structure transition of atomic carbon, 
were found redshifted to 87-103~GHz in a wide spectral scan\cite{strandet17}. The frequencies and spacings of these lines 
unambiguously place the galaxy at $z=6.900(2)$, corresponding to a cosmic age of 780~Myr (using recent cosmological 
parameters\cite{planck16cosmo} of Hubble constant $H_0$ = 67.7 km/s/Mpc, matter density $\Omega_{\rm m}$ = 0.309, and vacuum energy density $\Omega_\Lambda$ = 0.691). 
An elongated faint object is seen at optical and near-infrared wavelengths, consistent with a nearly edge-on spiral galaxy at $z=1.4\pm0.4$ 
acting as a gravitational lens for the background source (see Methods section `Modelling the SED'; here and elsewhere the error range quoted corresponds to a 1$\sigma$ uncertainty). 
Together these observations indicate that \sname\ is the most distant known member of the population of massive, 
infrared-bright but optically dim dusty galaxies identified with ground- and space-based wide-field surveys\cite{casey14}.

\begin{figure}
\includegraphics[width=\textwidth]{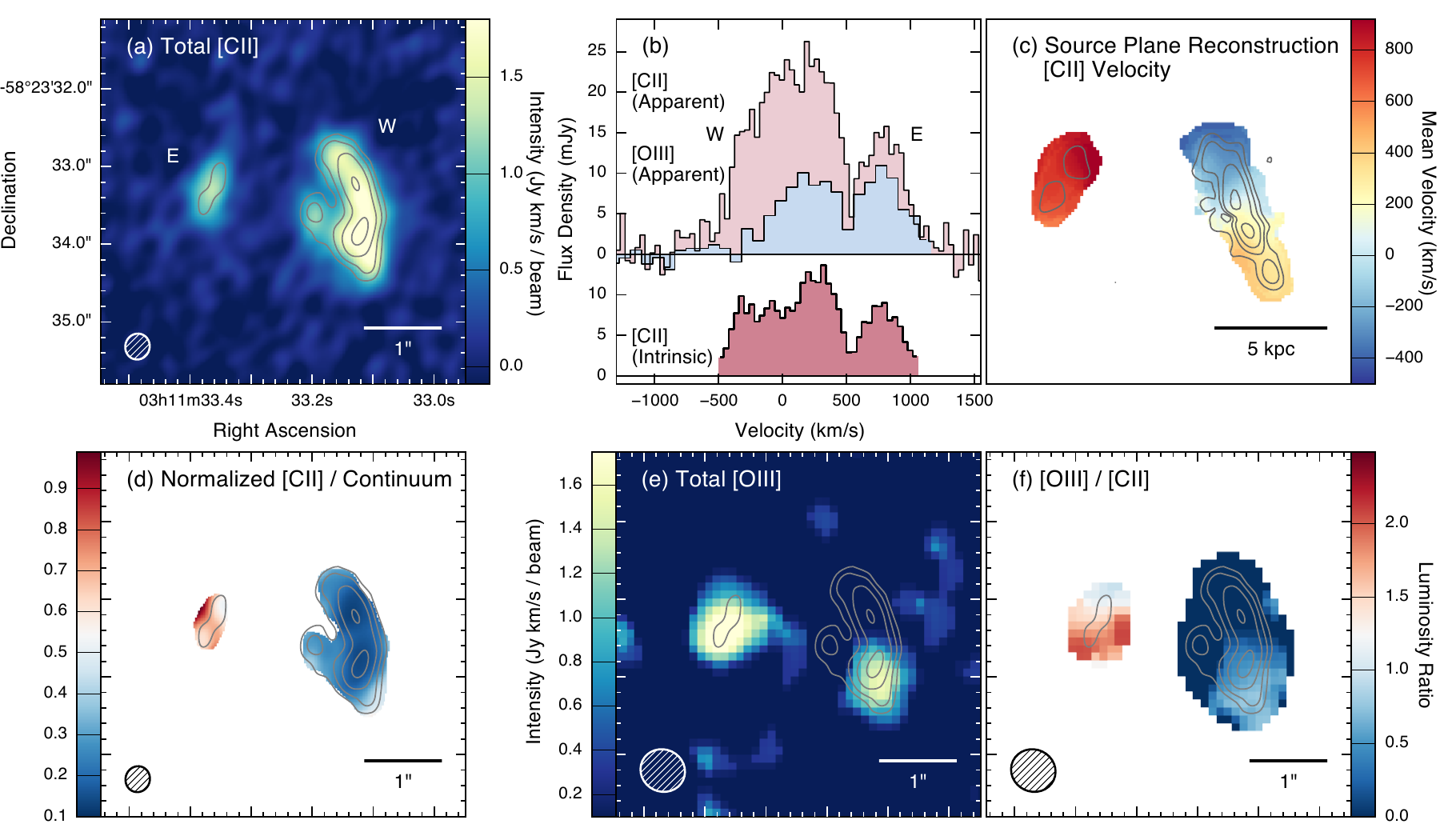}
\caption{\small
\textbf{Continuum, \cii, and \oiii\ emission from \sname\ and the inferred source-plane structure.} 
\label{f:image}
(a) Emission in the 157.74~\um\ fine structure line of ionized carbon (\cii) as measured at 240.57~GHz with ALMA, 
integrated across 1500~\kms\ of velocity, is shown with the color scale. The range in flux per synthesized beam 
(the 0.25$\times$0.30\arcsec\ beam is shown in the lower left), is provided at right. The rest-frame 160~\um\ continuum emission, 
measured simultaneously, is overlaid with contours at 8, 16, 32, 64 times the noise level of 34~$\mu$Jy~beam$^{-1}$. \sname E and W are labeled.
(b) The continuum-subtracted, source-integrated \cii\ and \oiii\ spectra. The upper spectra are 
as observed (``apparent'') with no correction for lensing, while the lensing-corrected (``intrinsic'') \cii\ spectrum is shown at bottom.
The E and W sources separate almost completely at a velocity of 500~\kms. 
(c) The source-plane structure after removing the effect of gravitational lensing. 
The image is colored by the flux-weighted mean velocity, showing clearly that the two objects are physically associated but 
separated by roughly 700~\kms\ in velocity and 8~kpc (projected) in space. The reconstructed 160~\um\ continuum emission is 
shown in contours. A scale bar in the lower right represents the angular size of 5~kpc in the source plane. 
(d) The line-to-continuum ratio at the 158~\um\ wavelength of \cii, normalized to the map peak. 
The \cii\ emission from \sname{E} is significantly brighter relative to its continuum than for W. The sky coordinates and rest-frame 160~\um\ continuum contours of Fig.~\ref{f:image}(d), (e), and (f) are the same as in panel (a).
(e) Velocity-integrated emission in the 88.36~\um\ fine structure line of doubly-ionized oxygen (\oiii) as measured at 429.49~GHz with ALMA.
The data have an intrinsic angular resolution of 0.2$\times$0.3\arcsec\ but have been tapered to 0.5\arcsec\ owing to the lower 
signal-to-noise in these data. 
(f) The luminosity ratio between the \oiii\ and \cii\ lines. As in the case of the \cii\ line-to-continuum ratio, a significant disparity 
is seen between the E and W galaxies of \sname. 
}
\end{figure}

The far-infrared emission from \sname\ provides an opportunity to study its structure with little confusion from the foreground galaxy.
We conducted 
ALMA observations at $\sim$0.3\arcsec\ resolution at three different frequencies (see \SI), 240, 350, and 
420~GHz, corresponding to rest-frame 160, 110, and 90~\um. The observations at 240 GHz include the 158~\um\ fine-structure 
line of ionized carbon (\cii) and those at 420 GHz the 88~\um\ fine-structure line of doubly ionized oxygen (\oiii). 
The 160~\um\ continuum and \cii\ and \oiii\ line emission maps of the source are shown in Fig~\ref{f:image}. 
Two emissive structures are visible in the map, {\sname}E and {\sname}W, 
which are separated by less than 2\arcsec\ on the sky, before correction for gravitational deflection. While the morphology of the eastern 
and western sources is reminiscent of a lensing arc (W) and counter-image (E), the \cii\ line clarifies the physical situation. {\sname}E 
is separated from the brighter W source by 700~\kms\ and is therefore a distinct galaxy.

Lens modeling of the 160, 110, and 90~\um\ continuum emission from this source was performed using a 
pixellated reconstruction technique\cite{hezaveh16} (Fig.~\ref{f:image}c, \EF~\ref{f:lenscont}, and \SI, Section~`Gravitational lens modelling'). 
The source structure and lensing geometry is consistent between the observations, 
and indicates that the two galaxies are separated by a projected (proper) distance of 8~kpc in the source plane. 
{\sname}E has an effective radius of 1.1~kpc, while {\sname}W shows a clumpy, elongated structure 
7.5~kpc across. The (flux-weighted) source-averaged magnifications of each galaxy and of the system as 
a whole are quite low ($\mu_{\rm E} = \muE$, $\mu_{\rm W} = \muW$, $\mu_{\rm tot} = \muTotCont$) because 
the W source is extended relative to the lensing caustic and the E source is far from the region of 
high magnification. The same lensing model applied to the channelized \cii\ data reveals a clear velocity 
gradient across {\sname}W which may be due to either rotational motions or a more complicated source structure 
coalescing at the end of a merger.

With the lensing geometry characterized, it is clear that the two galaxies that comprise \sname\ are extremely luminous. 
Their intrinsic infrared (8-1000~\um) luminosities have been determined from observations of rest-frame ultraviolet to 
submillimeter emission (see \SI, Section~\ref{s:SEDmodel}), and are $L_{\rm IR}=\LirE$ and 
$\LirW$~\lsol\ for E and W, respectively. Assuming that these sources are powered by 
star formation, as suggested by their extended far-infrared emission, these luminosities are unprecedented at $z>6$. 
The implied (magnification-corrected) star formation rates are correspondingly enormous, \sfrE\ and \sfrW~\msol~yr$^{-1}$, likely 
due to the increased instability associated with the tidal forces experienced by merging galaxies\cite{mihos96}.  
The components of \sname\ have luminosities and star formation rates similar to the other massive $z>6$ galaxies identified by their dust emission, 
including HFLS3 ($z=6.34$), 
which has a star formation rate of 1300~\msol~yr$^{-1}$ after correcting 
for a magnification factor\cite{cooray14} of 2.2, and a close quasar-galaxy pair resolved recently with ALMA\cite{decarli17} 
at $z=6.59$, which are forming stars at rates of 1900 and 800~\msol~yr$^{-1}$. Unlike the latter case, however, there is 
no evidence of a black hole in either source in \sname. 

Unlike any other massive dusty source at $z>6$, the rest-frame ultraviolet emission of \sname{E} is clearly detectable with modest 
integration by the \textit{Hubble Space Telescope}. 
The detected ultraviolet luminosity ($L_{\rm UV} = 7.4\pm0.7\times10^{10}$~\lsol) suggests a star formation rate of just 13~\msol~yr$^{-1}$, 2\% of the 
rate derived from the far-infrared emission, 
which is consistent with \sname{E} forming most of its stars behind an obscuring veil of dust. 
The inferred stellar mass for this galaxy (see \SI\ Sec.~\ref{s:SEDmodel}) is  
\mstarE~\msol. 
While no stellar light is convincingly seen from \sname{W}, the absence of rest-frame ultraviolet emission  
is likely explained by heavy dust obscuration and is not unusual\cite{casey17}. 
Although \sname{E} is the less massive of the two components, even this galaxy is rare among 
ultraviolet-detected galaxies at $z\sim7$. Such galaxies are found in blank field surveys to have a 
sky density of just 1 per 30 square arcminutes\cite{bouwens15}. 

The far-infrared continuum and line emission of \sname{E} and W seen in Fig.~\ref{f:image}(d-f) 
imply substantial differences in the physical conditions in these objects. 
Compared to the western source, \sname{E} 
has a higher ratio of \cii\ line to 160~\um\ continuum and a much larger luminosity ratio between \oiii\ and \cii.
The \oiii\ emission is dramatically more luminous in the eastern source, with most of the western source (excluding 
the southern end) showing no emission at all. Because the formation of O$^{++}$ ions requires photons with energies of more than 35.1~eV, 
this line arises only in ionized regions around the hottest stars or near active galactic nuclei\cite{ferkinhoff10}.
It is unlikely that active galactic nuclei are the origin for the \oiii\ line in \sname{E}, because both the continuum and line emission extend 
across most of the galaxy rather than being concentrated in a putative nuclear region. Observations of \oiii~88~\um\ 
emission in actively star forming galaxies at low\cite{brauher08,cormier15} and high\cite{inoue16} redshift have found that 
the line luminosity ratio between \oiii\ and \cii\ rises as gas metallicity decreases. The ultraviolet photons capable of forming O$^{++}$ have a longer mean free 
path in a lower-metallicity interstellar medium than in a higher-metallicity one, and the electron temperature remains higher for the same ionizing flux, both of which 
favor increased \oiii\ emission\cite{lebouteiller12}. 
The difference in the \cii\ line-to-continuum ratio may result from multiple effects: the known suppression\cite{diazsantos13,oteo16,spilker16} 
of the \cii\-to-\lir\ ratio in regions of increased star formation surface density (higher in \sname{W}), and the 
increased \cii\-to-\lir\ ratio in star-forming galaxies of lower metallicity\cite{cormier15}. Whether \sname{E} (or the southern edge
of \sname{W}, which is similar to E in these properties) has a more primordial 
interstellar medium than the bulk of \sname{W} can be tested in future observations.

\begin{figure}
\center\includegraphics[height=0.7\textheight]{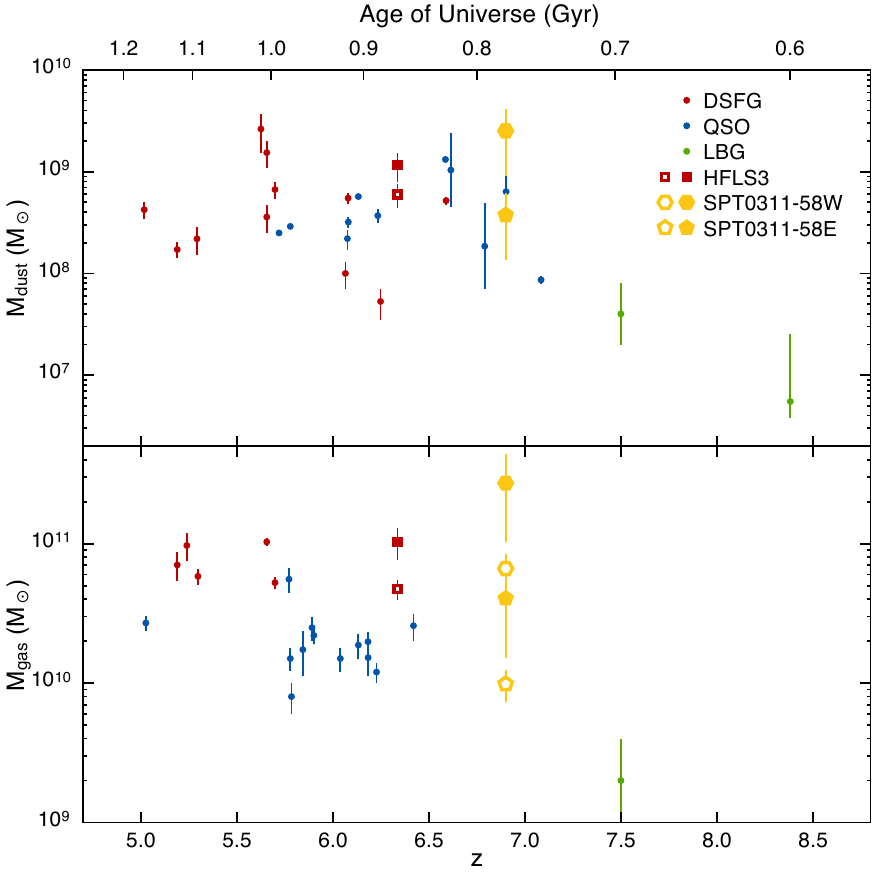}
\caption{\small
\textbf{Mass measurements for high-redshift galaxies.} 
\label{f:mass}
Dust masses are taken from the literature, as described in the \SI.
Gas masses are primarily derived from observations of various CO rotational transitions, 
with the literature line luminosities converted to molecular gas masses under standardized assumptions (see \SI).
The comparison sample is divided among three object classes: DSFGs, quasars (QSO), and Lyman-break galaxies (LBG).
These objects are typically selected by FIR emission (DSFGs) or optical/infrared emission (QSOs, LBGs). 
Three additional DSFGs, \sname{E} (yellow pentagons), \sname{W} (yellow hexagons) and HFLS3 (red squares), have extensive photometry and line measurements, which enable 
more sophisticated estimates of their dust and gas masses\cite{weiss07,strandet17} from a combined analysis of 
the dust and CO line emission. For these objects we also show masses derived under the simpler analysis as open symbols 
(for \sname\ the methods give very similar answers for \mdust). Error bars represent 1$\sigma$ uncertainties.
}
\end{figure}

The masses of the components of {\sname} are remarkable for a point just 780~Myr after the Big Bang. 
Fig.~\ref{f:mass} compares \sname\ to objects at $z>5$ for which we have estimates of dust mass (\mdust) and/or 
total gas mass (\mgas). For \sname\ the best constraints on both of these quantities come from the 
joint analysis\cite{strandet17} of its far-infrared continuum and line emission, specifically the rotational transitions of carbon monoxide and neutral carbon. 
Here we have divided these masses among the two galaxies according to the lensing-corrected 
ratio of dust continuum emission (6.7) observed in our three high-resolution ALMA continuum observations, as 
the dust continuum luminosity is roughly proportional to the dust mass. 
The corresponding dust and gas masses are $M_{\rm gas} = \mgasAXELw$ (W) and \mgasAXELe~\msol\ (E),
and $M_{\rm dust} = \mdustAXELw$ (W) and \mdustAXELe~\msol\ (E).
The gas mass can also be estimated using the CO luminosity, though the conversion between luminosity and 
gas mass in this optically thick line is well-known to vary substantially depending on many factors, including star formation 
intensity and metallicity\cite{bolatto13}. Taking the observed\cite{strandet17} luminosity in the $J=3-2$ line of CO, 
converting to $J=1-0$ under the conservative assumption of 
thermalized emission, and connecting luminosity to mass with a standard value of 
$\alpha_{\rm CO} = 1.0\,\msol \mathrm{(K\,km\,s^{-1}\,pc^2)}^{-1}$, we derive
$M_{\rm gas} = \mgasCOw$ (W) and $M_{\rm gas} = \mgasCOe$~\msol\ (E). 
\sname{W} stands out above all of the known galaxies at $z>6$, that is, during the first $\sim$900~Myr of cosmic history. 

\begin{figure}
\includegraphics[width=6in]{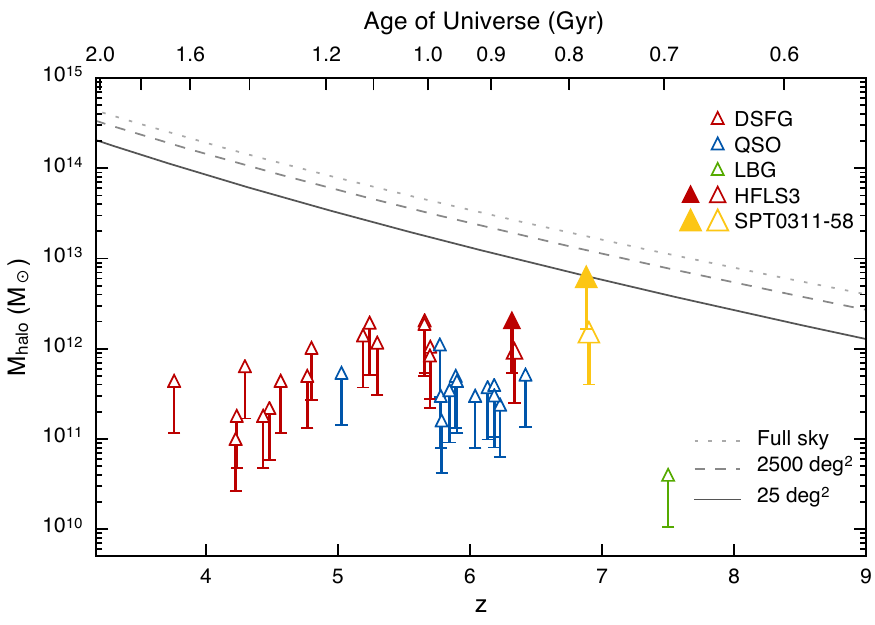} 
\caption{\small
\textbf{Halo masses for rare, high-redshift massive galaxies.} 
\label{f:rarity}
The dark matter halo mass ($M_{\rm halo}$; defined at an overdensity of 200 times the mean matter density of the universe) is inferred for galaxies in the first 2~Gyr after the Big Bang
(see \SI). These masses present a range of lower limits, from the most conservative assumption (lower bars) 
that all baryons in the initial halo have been accounted for in the molecular gas mass, 
to the observationally motivated assumption (upper triangles) 
that the baryonic mass ($M_{\rm b}$) in gas is a fixed ratio of the halo mass $M_{\rm b}/M_{\rm halo} = 0.05$, 
calibrated through a comparison\cite{behroozi13} of simulations and observations spanning $z=0-8$. 
The most massive halos that are expected to be observable\cite{harrison13} within the whole sky (dotted line), 
the 2500~deg$^2$ area of the SPT survey (dashed line), 
and within the subset of that area that is magnified by a factor of two or more (solid line) are also plotted as a function of redshift.  
As \sname{E} and W reside within the same halo, they are combined for this analysis. 
As in Fig.~\ref{f:mass}, halo masses are derived for HFLS3 (large red triangles) and \sname\ (large yellow triangles) using only the CO luminosity (open symbols) and the more 
sophisticated dust and CO analysis (filled symbols); the pairs of points are slightly offset in redshift for clarity.
}
\end{figure}

\sname\ highlights an early and extreme peak in the cosmic density field and presents an opportunity to test 
the predictions for the growth of structure in the current cosmological model.
The mass of the dark matter halo that hosts this system is uncertain, but it can be estimated in several ways. 
Considering the gas masses of the two galaxies, 
which, for most massive star forming galaxies\cite{bothwell13,aravena16}, represent the dominant component of baryons that have cooled and assembled at the center of the dark matter halo, the cosmic baryon 
fraction\cite{planck16cosmo} $f_{\rm b}=0.19$ places a hard lower bound on the total halo mass of $4\times10^{11}$~\msol\ 
for the lower ($\alpha_{\rm CO}$-based) gas mass estimate. 
A less conservative assumption incorporates the knowledge, based on observations across a wide range of redshifts, that only a 
fraction of the baryons in a dark-matter halo (less than one-quarter, $M_{\rm b}/M_{\rm halo} = 0.05$; see figure~15 of Ref.~\citenum{behroozi13}) 
are destined to accrue to 
the stellar mass of the central galaxy\cite{behroozi13}. In this case, a total halo mass of $(1.4-7.0)\times10^{12}$~\msol\ is implied, 
depending on which estimate of gas mass is adopted. 
To understand the rareness of the dark matter halo hosting \sname, we calculate curves describing the rarest halos that should exist in the 
universe at any redshift\cite{harrison13}. In Fig.~\ref{f:rarity} we show the halo masses inferred for many high-redshift galaxies, 
using the same path of gas mass to halo mass described above, finding that \sname\ is indeed closest to the exclusion curves and 
therefore marks an exceptional peak in the cosmic density field at this time.

We have found a system of massive, rapidly star forming, dusty galaxies at $z=6.900$, the most distant galaxies of this type yet discovered.
Two compact and infrared-luminous galaxies are seen separated by less than 8~kpc in projection, and 
700~\kms\ in velocity, probably in the process of forming one of the most massive galaxies of the era. Even before coalescence, 
the larger galaxy in the pair is 
more massive than any other galaxy known at $z>6$. 
Although the discovery of such a system at this high redshift and in a survey that covered less than 10\% of the sky is unprecedented, 
its existence is not precluded by the current cosmological paradigm.

\clearpage

\noindent\textbf{\Large{Methods}}

\section{ALMA Millimeter/Submillimeter Interferometry}
\label{s:almaSI}
We have acquired four observations of \sname\ with ALMA in four receiver bands (B3, B6, B7, B8, covering 84 to 432 GHz) 
under projects 2015.1.00504.S (PI: Strandet) and 2016.1.01293.S (PI: Marrone). A summary of these observations, including dates, 
calibration sources, integration times, atmospheric opacity, noise levels, and resolution, is provided in Extended Data Table~\ref{t:ALMA}. 
Salient details are provided below for each observation.

The redshift of \sname\ and the 3~mm continuum flux density were 
determined from an 84.2-114.9~GHz spectrum assembled from five separate tunings in ALMA band 3 under ALMA Cycle 3 
project 2015.1.00504.S. The observing strategy has been used to discover the redshifts of more than 50 SPT dusty 
sources, and further details on the redshift coverage are provided in previous works\cite{weiss13,strandet16}. 
Data were taken on 2015 December 28 and 2016 January 2 in ALMA configuration C36-1 (baseline lengths of 15-310~meters) 
using 34 and 41 antennas, respectively. The resulting image has a resolution of 3.3$\times$3.5\arcsec, though there is 
spatial information on somewhat finer scales that allows us to separately estimate flux densities for the E and W sources, separated by 
$\sim$2\arcsec. Further details of the analysis are provided by Strandet et al.\cite{strandet17}.

ALMA observed \sname\ a second time under 2015.1.00504.S in band 7 (LO = 343.48 GHz) to produce a continuum image 
suitable for gravitational lens modeling. Similar observations were used to produce lens models of SPT 
sources in previous cycles\cite{hezaveh13,spilker16}.  The observations were performed with 41 antennas in the C40-4 configuration, providing 
15--770~meter baselines. The resulting image has an angular resolution of 0.3$\times$0.5\arcsec, though lacking any spectral lines 
it was found to be insufficient to provide an unambiguous determination of the lensing configuration. 

The ALMA Cycle 4 project 2016.1.01293.S was intended to follow up on the discovery of this very distant source through spectroscopic 
observations. The 158~\um\ line of \cii\ was observed on 2016 November 3 in ALMA configuration C40-5, which provided 
baseline lengths of 18--1120~meters. This provides the primary imaging for this work, as it yielded an extremely significant detection of 
the \cii\ line and continuum structure at high resolution. 

A final observation was obtained in ALMA band 8 (LO = 423.63~GHz), in configuration C40-4 (baselines 15--920~meters). 
The observations were repeated in four segments to yield the required integration time. The resulting data have 0.2$\times$0.3\arcsec\
resolution. These data provide a final spatially-resolved continuum observation, at 90~\um\ rest-frame wavelength, along 
with spectroscopic images of the 88~\um\ line of \oiii. The ALMA continuum images 
are shown in \EF~\ref{f:ALMAall}.

\section{\textit{Spitzer} Infrared Imaging}
Infrared observations of \sname\ were acquired with the Infrared Array Camera (IRAC) instrument\cite{fazio04} on the 
\textit{Spitzer Space Telescope} as a part of 
Cycle 24 \textit{Hubble Space Telescope} program 14740 (PI: Marrone).
The observations consisted of 95 dithered 100\,s exposures on-source in both operable 
IRAC arrays at 3.6 and 4.5~\um.  A large dither throw was used.  The dataset thus had sufficiently high 
redundancy to support our standard reduction procedure, which involved constructing an object-masked 
median stack of all 95 exposures in each band, and then subtracting the median stack from the raw frames to 
compensate for bad pixels not automatically masked by the pipeline, and remove gradients in the background. 
After these initial preparatory steps, the background-subtracted exposures were combined in the 
standard way\cite{ashby13} with IRACproc\cite{schuster06} and MOPEX to create mosaics having 0.6\arcsec\ pixels.  
The mosaics achieved an effective total integration time of about 9000\,s after masking cosmic rays and other 
artifacts.  Two flanking fields were covered to the same depth but separately, each in one IRAC passband.

Photometry was performed on the mosaics using Source Extractor\cite{bertin96} (SE) in dual-image mode 
after trimming to exclude the flanking fields and unexposed areas.  The lens galaxy associated with \sname\ was well detected 
with no evidence for saturation or even nonlinear detector behavior.   During this process background and object 
images were generated and inspected to verify that SE performed as expected and generated valid photometry.

\section{\textit{Hubble Space Telescope} Imaging}
\sname\ was observed for 5 orbits of \textit{HST} imaging with ACS and WFC3/IR in Cycle 24 (PID 14740; PI: Marrone) 
to determine the morphology of the foreground lens and better constrain the spectral energy distribution 
of both the lens and source. All observations were acquired on 30 April 2017.  The ACS imaging consists of a single 
orbit divided between the F606W and F775W filters. Exposure times are 844 s and 1.5 ks, respectively.  
Four orbits of WFC3/IR observing was split evenly between the F125W and F160W filters. While the nominal exposure 
times are 5.6 ks, a subset of the data in both filters was compromised by significant contamination 
from scattered earthlight. We reprocessed the imaging to remove contaminated data, resulting in final 
exposure times of 4.9ks in each band.

\section{Gemini Optical/IR Imaging and Spectroscopy}
With the Gemini Multi-Object Spectrograph\cite{hook04} (GMOS) of Gemini-South, we obtained deep
$i$ and $z$ images of \sname\ (Program: GS-2015B-Q-51, PI: Rotermund) on 2016 January 29 and 31.
The instrument consists of three 2048 x 4176 pixel CCDs, separated by two
6.46\arcsec\ (80 pixel) gaps, with a scale of 0.0807\arcsec~pixel$^{-1}$. The field of view (FOV) of the GMOS camera is 
5.5\arcmin$\times$5.5\arcmin.
Our images were taken under photometric conditions and using a 2$\times$2 binning,  which gives a scale of 0.161\arcsec~pixel$^{-1}$.
The total integration times were 3600~s for $i$-band and 6600~s for $z$-band, with average seeing conditions of 
1.3\arcsec\ and 1.0\arcsec\ in $i$-band and $z$-band, respectively. The resulting 5$\sigma$ point source depths 
were $i_{AB} =25.2$ and $z_{AB} =25.0$.

\sname\ was observed using the Facility Near-Infrared Wide-field Imager \& Multi-Object Spectrograph for Gemini (FLAMINGOS-2)\cite{eikenberry12} at the Gemini-South Observatory on the nights of 
UT 2016 September 23, and 2017 February 06, under Program GS-2016B-Q-68. 
The instrument was used in imaging mode, with 0.181\arcsec\ pixels, and yielding an unvignetted circular field of view of $\sim$5.5\arcmin\ 
diameter. Our observing sequence for the survey consisted of a randomly ordered dither pattern, 
with 15\arcsec\ offsets about the pointing center. This pattern was repeated until the
required total exposure time was achieved. The individual K$_s$-band exposure time was set at 15~s 
in the first observation, and 10~s in the second observation, yielding a typical background sky level in 
$K_s$ of $\sim$10,000-12,000 counts (detector nonlinearity can be corrected to better than 
1\% up to 45,000 counts), ensuring that 2MASS stars with $K_s>13$ do not saturate and can be 
used for photometric calibration. 
The data were reduced using the python-based FLAMINGOS-2 Data Pipeline, FATBOY\cite{warner12,warner13}, created at the University of Florida. 
Briefly, a calibration dark was subtracted from the data set, a flat field image and a bad pixel map were 
created, and the flat field was divided through the data. Sky subtraction was performed to remove small-scale 
structure with a subsequent low order correction for the large-scale structure. Finally, the data were aligned and 
stacked. The seeing conditions averaged 0.7\arcsec\ in the final image comprising 44 minutes of integration, 
reaching $K_{s,AB}=23.6$ at 5$\sigma$.

Spectroscopy was obtained with the GMOS-S instrument on the nights of UT 2016 February 1 and 2 
(Program: GS-2016B-Q-68, PI: Chapman) using 
the 1\arcsec\ wide longslit at a position angle~$=-10^\circ$ east of north, and the instrument configured 
with the R400 grating and 2$\times$2 detector binning. For a source that fills the 1\arcsec\ slit this setup 
results in a spectral resolution of $\sim7$\AA. The observations were spectrally dithered, using two central wavelength 
settings (8300 and 8400\AA) to cover the chip gaps. The data comprise a series of individual 900s exposures, 
dithering the source spatially between two positions (``A'' and ``B'') along the slit in an ABBA pattern, repeated 
four times, two at each central wavelength setting. The total integration 
time is 4 hrs. A bright foreground object was positioned along the slit midway between the acquisition star 
and \sname, providing an additional reference point for locating traces along the slit.

The spectra were reduced, beginning with bias subtraction and bad pixel masking using the  {\sc IRAF} 
GMOS package provided by Gemini. The individual chips were combined into a single mosaic for each exposure and 
the mosaicked frames were then sky subtracted by differencing neighboring A--B exposure pairs; 
this method resulted in nearly Poisson noise even under the numerous bright sky lines. A 
flat-field slit illumination correction was applied and a wavelength calibration derived for each mosaic. 
The 2-dimensional spectrum was created by median combining the individual exposure frames.

The spectrum shows a faint continuum beginning above 9000\AA\ at the location of \sname. A one-dimensional 
extraction of the faint trace yields no reliable redshift measurement, but it is consistent with the redshifted 
4000\AA\ break that is expected for the foreground galaxy at $z\sim1.4$. Calibrated against the nearby $R$=16.4 star spectrum we find no 
flux at the expected location of Ly$\alpha$ redshifted to $z=6.900$ ($\sim$9600\AA) down to a 3$\sigma$ 
flux limit of $3.0\times10^{-17}$ erg s$^{-1}$ cm$^{-2}$ for a 500~\kms\ wide emission line.

\section{Image de-blending}
\label{s:deblend}
At the position of \sname, our optical and infrared images (\EF~\ref{f:oir}) show a prominent lower-redshift galaxy responsible for lensing 
the western source, as well as direct stellar emission from the eastern source in the \hubble\ images (\EF~\ref{f:3color}), which have the highest resolution. 
To extract reliable photometry for \sname{E}, particularly in the low-resolution \spitzer\ images that 
cover the rest-frame optical, and to search for emission from W underneath the lens galaxy, we must model and 
remove the lens emission. We follow procedures similar to those in past work\cite{ma15}, using the \hst/WFC3 images as 
the source of the lens galaxy model to de-blend the IRAC image. 
The foreground lens can be fit with a single S\'ersic profile with an index $n=1.77$.
As seen in \EF~\ref{f:deblend}, there is no clear rest-frame ultraviolet emission from \sname{W} in the \hst\ bands after removal of the lens model. 
To remove the lens from the IRAC image, the WFC3 model is convolved with the IRAC point spread function and 
then subtracted from the 3.6 and 4.5~\um\ images. Residual emission is seen near the positions of both the eastern and western sources. 
Unfortunately, because \sname{W} lies right on top of the lens, the residuals are extremely susceptible to image deconvolution errors 
and we do not believe the \spitzer/IRAC fluxes to be reliable. In contrast \sname{E} is one full IRAC resolution element, 1.7\arcsec, 
from the lens centroid, and we consider the residual emission at this position to be usable in our subsequent analyses.
Images of the model and residuals are provided in \EF~\ref{f:deblend} and the resulting photometry is provided in Extended Data Table~\ref{t:OIRdata}.

\section{Gravitational Lens Modeling}
\label{s:lensmodel}
Gravitational lens modeling of \sname\ was performed using two different codes that model the source-plane emission in different ways. Both codes fit directly to the visibilities measured by ALMA or other interferometers to avoid the correlated noise between pixels in inverted images. In each, the lens galaxy is modeled as a singular isothermal ellipsoid, and posterior parameter distributions are sampled using a Markov Chain Monte Carlo technique, marginalizing over several sources of residual calibration uncertainty (e.g., antenna-based phase errors). 

Initial lens models were created using the \texttt{visilens} code, which is described in detail elsewhere\cite{spilker16}. The source plane is modeled as one or more elliptical S\'{e}rsic profiles. Because of the simplicity of this source-plane representation, the code is able to sample large and complex parameter spaces quickly. The continuum emission at 160, 110, and 90~\um\ was modeled with four S\'{e}rsic components, one for the {\sname}E and three for W. These models leave $\sim$8$\sigma$ peak residuals in the 160 and 90~\um\ data, which both reach peak signal-to-noise $>$150.

After determining the lens parameters using \texttt{visilens}, we used the best-fit values as initial input to a pixellated 
reconstruction code\cite{hezaveh16}.  This code represents the source plane as an array of pixels, rather than an analytic model, 
and determines the most probable pixel intensity values for each trial lens model while imposing a gradient-type 
regularization\cite{suyu06} to avoid over-fitting the data. For each dataset, we fit for the strength of this regularization. At 160 and 90\,\um\, we re-fit for the lens model parameters and compare to the \texttt{visilens} models as a test of the robustness of the lens modeling. Within each code, the best-fit lens parameters at the two independent wavelengths are consistent to $<$10\%. Further, both the lens parameters and source structure are consistent between the two independent codes, with intrinsic source flux densities, sizes, and magnifications that agree to $<$15\%.  The increased freedom in the source plane afforded by the pixellated reconstruction means that the lens parameters are not independently well-constrained by the 110\,\um\ data, which have both lower signal-to-noise and spatial resolution. For these data, we simply apply the lensing deflections determined from the other two datasets to reconstruct the source-plane emission. The pixellated reconstructions of the three continuum wavelengths are shown in \EF~\ref{f:lenscont}.

The channelized \cii\ line is also modeled using the same pixellated 
reconstruction technique, using 39 consecutive channels of 40~km/s width, each with a peak signal-to-noise ranging from 9--34. For each channel, we apply the lensing deflections from the best-fit model of the 160\,\um\ data, which were observed simultaneously. We fit for the strength of the source-plane regularization\cite{suyu06,hezaveh16} at each channel, which varies across the line profile as some velocities experience higher magnification (e.g., those multiply imaged from -280\,km/s to +80\,km/s) than others (e.g., the entire eastern source at $>$+560\,km/s). The models of each \cii\ channel are represented in \EF~\ref{f:lenscii}.

We determine the source magnifications using the 90~\um\ pixellated model, in which the E source is detected at the highest signal-to-noise ratio and so the effects of varying the aperture used to measure the intrinsic flux density are minimized. Because the source-plane morphology is very similar between the three continuum wavelengths, the magnification is also essentially identical between them. We find flux-weighted, source-averaged magnifications for the eastern, western, and the system as a whole of $\mu_{\rm E} = \muE$, $\mu_{\rm W} = \muW$, $\mu_{\rm tot} = \muTotCont$, respectively.  These magnifications are substantially lower than the median magnification of 5.5 within the sample of 47 SPT-discovered dusty galaxies\cite{spilker16} for which we have data adequate to construct lens models or conclude that sources are unlensed. In this case the low magnification is a consequence of the low mass of the  lensing halo, typically expressed as an ``Einstein'' radius, $\theta_{E}$. The lens model for this source indicates $\theta_{E}=0.29$\arcsec, which is around the 10$^{\rm th}$ percentile for SPT lensed sources\cite{spilker16}, and the background source is both much larger than and offset from the regions of highest magnification. A large portion of the source is therefore only weakly magnified and the source-averaged values are low.

Finally, we also construct a lens model of the 95~GHz ALMA data (rest-frame 380~\um; Extended Data Table~\ref{t:ALMA}). Because the spatial resolution of these data is low (3.5\arcsec), we model these data using only the \texttt{visilens} code, which is more suited for low-resolution data. We only allow the lens parameters and source structural parameters (e.g., position, radius) to vary within the ranges determined from the higher-resolution 160, 110, and 90~\um\ continuum data, leaving only the flux densities of the eastern and western objects as free parameters. This modeling indicates that essentially all of the observed 380\,\um\ emission can be ascribed to the western source, with the eastern source ``detected'' at $\sim$1$\sigma$.

In addition to the ALMA data, we use \textit{Herschel} photometry reported previously\cite{strandet17} to constrain the SED of 
{\sname}E and W to rest-frame 30~\um\ (250~\um\ observed). The resolution of \textit{Herschel} SPIRE is not adequate 
to separate the two components, so we divide the total flux density observed in the three SPIRE bands between E and W 
according to the ratios observed in the ALMA bands. These photometric points are then corrected for the continuum magnification 
derived from the ALMA data and used in the SED modeling described below. The 
total and intrinsic flux densities are reported in Extended Data Table~\ref{t:FIRdata}.

\section{Modeling the Spectral Energy Distribution}
\label{s:SEDmodel}

\EF~\ref{f:fullSED} presents the spectral energy distributions of \sname{E}, W, and the foreground lens galaxy.

A photometric redshift for the lens is calculated with EAZY\cite{brammer08} using the data in Extended Data Table~\ref{t:OIRdata}. 
The resulting redshift is 1.43, with a 1$\sigma$ confidence interval of 1.08$-$1.85. 
The lens SED fitting is performed with the Code Investigating GALaxy Emission (CIGALE\cite{burgarella05,noll09}) 
assuming $z=1.43$. 

The multiple rest-frame ultraviolet to rest-frame optical detections of \sname{E} allow us to constrain the stellar mass using reasonable 
assumptions about the star formation history at this early point in cosmic history. The SED is fit by varying the 
e-folding time and age from a previously reported stellar population model\cite{bruzal03} under  
single- and two-component formation histories, assuming solar metallicity and using a previously reported\cite{chabrier03} initial mass function. The minimum radiation field, power-law slope, and gamma, the fraction of dust mass exposed to radiation intensities above the minimum, from one dust model\cite{draine07}, and the color excess and attenuation slope  
from other dust models\cite{calzetti00,leitherer02} are kept free in the SED fitting. 
The AGN contribution is set to zero as there are no photometric points to constrain the spectral range most affected by AGN power 
(mid-infrared) and thus any fraction between 0 and 60\% of the dust luminosity is attributable to AGN with nearly equal probability. However, 
this ignores the spatial distributions of the dust and line emission, which are not strongly peaked as is usually observed in AGN-dominated galaxies, 
so we deem this wide range to be unphysical. 
The inferred stellar mass and star formation rates are \mstarE~\msol\ and 
\sfrE~\msol~yr$^{-1}$, respectively, for the two-component star formation history. These answers agree within the uncertainties for 
a single-component star formation history. 
The infrared luminosity (\lir, integrated from 8--1000~\um) is \LirE~\lsol\ and the extinction is $A_V = 2.7\pm0.2$ mag. 

For the western source, we have only upper limits and the potentially-contaminated IRAC detections to constrain the rest-frame optical/ultraviolet 
emission. Accordingly, we use the IRAC photometry as upper limits, along with the \hst\ limits and far-infrared data in Extended Data Table~\ref{t:FIRdata}, 
and model the SED with CIGALE. We find a luminosity of $\lir=\LirW$~\lsol, seven times larger than for the 
eastern source. A consistent luminosity is obtained by fitting the FIR SED with a modified blackbody\cite{blain03}.
The inferred star formation rate, 
which is closely connected to \lir, is \sfrW~\msol~yr$^{-1}$. 
Just as for the eastern source, the SED allows the AGN fraction to fall between 0 and 60\% at roughly equal probability, so we take 
the absence of a dominant IR emission region (see Fig.s~\ref{f:image}(c) and \ref{f:lenscont}) as an indication that the AGN contribution 
is unlikely to be important and fix the AGN fraction to zero. The dust luminosity due to star formation could therefore in 
principle be up to a factor of two smaller if the spatial distribution of the emission is ignored. 
Given that the photometry reaches only to rest-frame V band, it remains possible to 
hide a very large stellar mass behind plausible values for the extinction 
($A_{\rm V} \leq6$, as seen in other massive dusty galaxies\cite{hopwood11,lofaro13,cooray14,casey17}). 
Considering the IRAC flux densities alone, we can calculate simple rest-frame mass-to-light ratios for the observed bands to 
see what masses could exist without relying on the poorly-constrained CIGALE SED modeling. 
We use a stellar population synthesis code\cite{conroy09,conroy10} to compute a stellar mass-to-light ratio 
under a range of assumptions: stellar ages of 0.1--0.8~Gyr (from a reasonably ``young'' population to 
the approximate age of the universe at the time) and 
metallicity of 0.1--1~$Z_\odot$, with no dust attenuation. The mass then ranges from $(2-10)\times10^{10}~\msol$ per $\mu$Jy of 
measured flux density. Taking the measured and demagnified flux density (averaged between the two wavelengths) 
of 0.5~$\mu$Jy, we find a stellar mass of $(1-5)\times10^{10}$~\msol\ before correcting for extinction. 
If the extinction is as large as 5 magnitudes, the true stellar mass could be unphysically large ($>10^{12}$~\msol), 
demonstrating that we have no useful constraint without greater certainty about the reliability of the IRAC flux densities and/or more 
photometric data points.


\section{Galaxy and Halo Masses}

Figures~\ref{f:mass} and \ref{f:rarity} compile mass measurements for 
high-redshift galaxies discovered by various techniques. The galaxy sample primarily comprises galaxies 
identified through their luminous dust emission 
(DSFGs) and optically-identified quasars, which are typically the objects with the largest gas, dust, or stellar masses 
at these redshifts. At the very highest redshifts, where very few galaxies have been found, 
objects selected based on their ultraviolet emission are also included. 
The subsets of galaxies included in each figure 
overlap significantly but are not identical because not all of the requisite information is available for each source. 

\textit{Dust Mass}: Mass estimates are literature values unmodified from the original publications\cite{wang11,wang11b,walter12,riechers13,watson15,venemans16,venemans17,laporte17,decarli17} 
owing to the heterogeneity of the data available across the sample. The dust masses are generally derived 
from the FIR continuum emission, using one to several wavelengths. Differences between the cosmology 
assumed in this work and the past publications result in unimportant corrections and are ignored.

\textit{Gas Mass}: Following standard observational practice, the primary source for gas masses\cite{walter03,riechers10,wang10,wang11,wang11b,walter12,wang13,riechers13,rawle14,aravena16,venemans17} 
shown in Fig.~\ref{f:mass}(b) is measurement of the luminosity of rotational transitions of 
CO. The lowest available rotational transition is typically used and any translation between the observed 
transition and the $J=1-0$ line that is most commonly used as a molecular gas indicator is taken 
from the original source. Rather than accepting the varying coefficients for the conversion of CO luminosity to 
gas mass, we re-calculate all masses using a common value of 
$\alpha_{\rm CO} = 1.0\,\msol \mathrm{(K\,km\,s^{-1}\,pc^2)}^{-1}$, 
which is a typical value for actively star-forming galaxies\cite{solomon05,carilli13}.
For one source the gas mass is estimated through the star-formation surface density\cite{watson15}.

\textit{Halo Mass}: The halo masses of Fig.~\ref{f:rarity} are derived from the gas mass sample above. 
Each dark matter halo mass is represented using a range of values, starting with a conservative and 
hard lower limit found by dividing the measured gas mass by the universal baryon 
fraction\cite{planck16cosmo}, $f_{\rm b} = 0.19$. This lower limit ignores 
any baryonic mass that has been converted into stars and/or hot or cool atomic gas phases, 
which would increase the inferred halo mass. A more realistic, but still conservative, 
lower limit is represented by the top of the plotted symbols. Here we assume that the ratio of baryonic 
mass to halo mass is $M_{\rm b}/M_{\rm halo} = 0.05$. This value is a factor of $\sim 4$ less than the 
universal baryon fraction but still higher than the typical stellar-to-halo mass ratio inferred for halos of any 
mass and redshift via subhalo abundance matching\cite{behroozi13}. Given that we do not expect
high-mass galaxies such as \sname\ to expel a large fraction of their molecular gas content\cite{hayward17} 
or to later accrete dark matter without also accreting gas in proportion to the universal baryon fraction, it is 
reasonable to expect that the baryon-to-halo mass ratio should be less than this inferred upper limit
on the stellar-to-halo mass ratio across all masses and redshifts.

\textit{HFLS3 and \sname\ Masses}: 
In the case of the two most distant DSFGs, HFLS3\cite{riechers13} and \sname, which have extensive FIR photometry and 
atomic/molecular line measurements, we also compute the gas mass using a joint continuum/line radiative transfer 
model described in Refs.~\citenum{weiss07} and \citenum{strandet17}. The mass for \sname\ was computed in Ref.~\citenum{strandet17} 
without spatially-resolved (CO and \ci) line emission. For the purposes of Fig.~\ref{f:rarity} only the total 
gas mass of the two \sname\ galaxies is important for estimating the halo mass. For Fig.~\ref{f:mass} the dust mass is 
divided between the two sources according to the ratio of dust continuum emission in our resolved observations. The gas 
mass is similarly divided, though the velocity profile of the CO lines provides weak evidence that the molecular gas is concentrated 
in \sname{W}, which would increase the gas mass for this source by 15\%. 

\section{Calculation of Halo Rareness}
Figure \ref{f:rarity} demonstrates the ``rareness'' of \sname\ by considering its position in the dark matter 
halo mass-redshift plane compared with other extreme high-redshift objects (DSFGs, quasars, and an LBG) 
believed to be hosted by massive dark matter halos. 
To quantify the rareness of these extreme objects we employ the method of Ref.~\citenum{harrison13}, 
adopting their {\sc matlab} script available at \url{https://bitbucket.org/itrharrison/hh13-cluster-rareness} and modified 
slightly to extend the calculation to $z = 10$. These authors describe how to properly
compute $(z,M_{\rm halo})$ contours (``exclusion curves'') above which the Poisson probability of such an object 
being detected in the standard $\Lambda$CDM cosmology is less than $\alpha$ ($\alpha < 1$); the existence of a single object 
above such an exclusion curve is sufficient to rule out $\Lambda$CDM at the $100(1-\alpha)$\% confidence level. 
In Fig.~\ref{f:rarity}, we plot 1$\sigma$ exclusion curves (i.e., $\alpha = 0.32$).
Of the three different statistical measures of rareness that they propose, we employ the `$>\nu$' measure, 
which quantifies the rareness according to the minimum height of the primordial density perturbation from which a halo 
of mass $M_{\rm halo}$ and redshift $z$ could have formed, $\nu(M_{\rm halo},z) \propto (D_+(z) \sigma(M_{\rm halo}))^{-1}$,
where $D_+(z)$ is the normalized linear growth function and $\sigma^2(M_{\rm halo})$ is the variance of the
matter power spectrum smoothed on the comoving spatial scale that corresponds to the mass $M_{\rm halo}$. This 
statistic is sensitive to changes in the $\Lambda$CDM initial conditions, such as primordial non-Gaussianity 
(which would lead to more high-mass dark matter halos at a given redshift than expected in the
standard $\Lambda$CDM cosmology). 
For the purposes of this calculation, we assume a $\Lambda$CDM cosmology with parameters\cite{planck16cosmo} 
$(\Omega_m = 0.309, \Omega_{\Lambda} = 0.691, h_0 = 0.677, \sigma_8 = 0.816)$  and employ 
the halo mass function of Ref.~\citenum{tinker08}.

The $>\nu$ rareness statistic (and the corresponding exclusion curves) depends on the region of the $M_{\rm halo}-z$ plane 
to which the survey is sensitive. We assume that the SPT sample of lensed DSFGs is complete for $z > 1.5$.
At lower redshift, the probability of lensing is strongly suppressed\cite{hezaveh11,weiss13}, which 
means that the galaxy (or galaxies) associated with a halo of $M_{\rm halo} \gtrsim\ 10^{15} \msol$ (the $M_{\rm halo}$ 
value of the exclusion curves for $z = 1.5$) would have to have a very high intrinsic (i.e., unlensed) mm flux 
density ($S_{1.4} \gtrsim 20$ mJy) to be included in the sample. Because of the effects of ``downsizing'' (i.e., star 
formation is terminated at higher redshift in higher-mass galaxies than in lower-mass galaxies), it is unlikely 
that massive galaxies at $z < 1.5$ would have sufficiently high IR luminosity to be detected by the SPT\cite{miller15}.
We furthermore assume that the survey is complete for $M_{\rm halo} > 10^{11} \msol$. The assumption that the 
sample is complete to $M_{\rm halo} = 10^{11} \msol$ is a conservative one because the galaxies hosted by such 
halos, which would have $M_{\rm b} \ll 10^{10} \msol$, are unlikely to be sufficiently luminous to be detected without being 
very strongly lensed ($\mu \gtrsim 10$); erring on the side of overestimating the completeness yields a lower limit on the 
rareness. Substituting a minimum halo mass of, e.g., $10^{12}$~\msol\ would make the 
value of the $>\nu$ rareness statistic less than found for $10^{11}$~\msol, i.e., \sname\ would be inferred to be even rarer.

The total area from which the SPT DSFG sample was selected is 2500 deg$^2$. However, the fact that most 
of the SPT DSFGs are strongly lensed implies that the effective survey area is potentially much less than 2500 deg$^2$ 
because not only must a galaxy have a high intrinsic mm flux density to be included in the sample but also it must be 
gravitationally lensed by some factor in order to exceed the $S_{\rm 1.4\,mm} \simeq 20$~mJy threshold for inclusion
in redshift followup observations.
Properly accounting for the effects 
of lensing on the sample completeness would require defining an effective survey area as a function of halo 
mass and redshift, $A_{\rm eff}(M_{\rm halo},z) = 2500~{\rm deg}^2 P(\mu_{\rm min} | M_{\rm halo}, z)$, where 
$P(\mu | M_{\rm halo}, z)$ is the probability of a galaxy hosted by a halo of mass $M_{\rm halo}$ at redshift $z$ being 
lensed by a factor $\mu_{\rm min}$, the minimum magnification necessary for which a halo of mass $M_{\rm halo}$ 
and redshift $z$ would be detectable. However, given the large uncertainties in determining such a function, we 
opt for a simpler approach. Instead, in Fig. \ref{f:rarity}, we plot exclusion curves for the full sky (dotted line); 
an area of 2500 deg$^2$ (dashed line), which corresponds to the assumption that all halos in the mass and redshift 
range specified above would be detected even if they were not lensed; and an area of 25 deg$^2$ (solid line), 
which corresponds to the assumption that the survey area corresponds to just the $\sim1$\% of the SPT fields over 
which the magnification for sources at $z\gtrsim1.5$ will be at least\cite{hezaveh11,weiss13} $\mu=2$, like \sname. 



\noindent
\textbf{Code Availability} 
The lensing reconstruction for the ALMA data was initially performed with the visilens code available at \url{https://github.com/jspilker/visilens}. 
Pixelated reconstructions were performed using a proprietary code developed by a subset of the authors and additional non-authors, and we 
opt not to release this code in connection with this work. The rareness calculation was performed with publicly available code 
(\url{https://bitbucket.org/itrharrison/hh13-cluster-rareness}). The image deblending for the \spitzer\ images used GALFIT
(\url{https://users.obs.carnegiescience.edu/peng/work/galfit/galfit.html}). The SED modeling used the CIGALE code (\url{https://cigale.lam.fr/}), 
version 0.11.0. The photometric redshift of the lens galaxy was estimated with EAZY (\url{https://github.com/gbrammer/eazy-photoz}).

\noindent
\textbf{Data Availability} 
This paper makes use of the following ALMA data: ADS/JAO.ALMA\#2016.1.01293.S and ADS/JAO.ALMA\#2015.1.00504.S, available
at http://almascience.org/aq?project{\textunderscore}code=2015.1.00504.S and 
http://almascience.org/aq?project{\textunderscore}code=2016.1.01293.S. 
The \hst\ data are available online at the Mikulski Archive for Space Telescopes (MAST; https://archive.stsci.edu) 
under proposal ID 14740. Datasets analysed here are available from the corresponding author on reasonable request. 

\bibliographystyle{naturemag}
\bibliography{../bibtex/spt_smg.bib}


\begin{addendum}
\item 
ALMA is a partnership of ESO (representing its member states), NSF (USA) and NINS (Japan), 
together with NRC (Canada) and NSC and ASIAA (Taiwan), in cooperation with the Republic of Chile. 
The Joint ALMA Observatory is operated by ESO, AUI/NRAO and NAOJ.
This work incorporates observations with the NASA/ESA \textit{Hubble} Space Telescope, obtained at the Space Telescope 
Science Institute (STScI) operated by AURA, Inc. This work is also based in part on observations 
made with the \spitzer\ Space Telescope, which is operated by the Jet Propulsion Laboratory, 
California Institute of Technology under a contract with NASA. 
The SPT is supported by the National Science Foundation through grant PLR-1248097, with partial support through 
PHY-1125897, the Kavli Foundation and the Gordon and Betty Moore Foundation grant GBMF 947. 
Supporting observations were obtained at the Gemini Observatory, which is operated by the Association of 
Universities for Research in Astronomy, Inc., under a cooperative agreement with the NSF on behalf of the 
Gemini partnership: the National Science Foundation (United States), the National Research Council (Canada), 
CONICYT (Chile), Ministerio de Ciencia, Tecnolog\'{i}a e Innovaci\'{o}n Productiva (Argentina), and 
Minist\'{e}rio da Ci\^{e}ncia, Tecnologia e Inova\c{c}\~{a}o (Brazil). 
D.P.M., J.S.S., J.D.V., K.C.L., and J.S. acknowledge support from the U.S. National Science Foundation under grant AST-1312950. 
D.P.M. was also partially supported by NASA through grant HST-GO-14740 from the Space Telescope Science Institute and 
K.C.L. was partially supported by SOSPA4-007 from the National Radio Astronomy Observatory.
The Flatiron Institute is supported by the Simons Foundation.
J.D.V. acknowledges support from an A. P. Sloan Foundation Fellowship.

 \item[Author Contributions] D.P.M. proposed the ALMA \cii\ and \oiii\ line observations and analyzed all ALMA data. J.S.S. performed 
 the lens modeling. C.C.H. led the rareness analysis. M.L.N.A., M.B.B., S.C.C, A.H.G, J.M., K.M.R., and B.S. provided optical/infrared data reduction and 
 deconvolution. K.A.P. and J.D.V. performed SED modeling of the sources and lens. A.W. performed joint dust/line modeling of high-redshift targets. 
 D.P.M. wrote the manuscript. J.S.S., C.C.H., D.P.M., S.L., K.A.P., and J.D.V. prepared the figures. 
 All authors discussed the results and provided comments on the paper. The authors are ordered alphabetically after J.D.V.  
 
\item[Author Information] Reprints and permissions information is available at www.nature.com/reprints.
The authors declare no competing financial interests. 
Readers are welcome to comment on the online version of the paper. 
PublisherÕs note: Springer Nature remains neutral with regard to jurisdictional claims in published maps and institutional affiliations. 
Correspondence and requests for materials
should be addressed to D.P.M.~(dmarrone@email.arizona.edu).
\end{addendum}


\clearpage

\renewcommand{\tablename}{Extended Data Table}
{\renewcommand{\familydefault}{\sfdefault}
\setcounter{table}{0}

\begin{table}
\fontfamily{phv}\selectfont
\centering
\caption{ALMA Observations
\label{t:ALMA}}
\scriptsize
\begin{tabular}{lcccllccc}
\hline
Date & Frequency$^\mathrm a$ & Antennas & Resolution & Flux  & Phase & PWV$^\mathrm b$& $\mathrm t_{\mathrm{int}}^\mathrm c$ & Noise Level$^\mathrm d$\\
 & (GHz) & & (arcsec) & Calibrator & Calibrator & (mm) & (min) & ($\mu$Jy/beam) \\
\hline
\textbf{B3} & & & 3.3$\times$3.5 & & & & & 35\\
2016-Jan-02 & 91.95 & 41 & 3.8$\times$3.9 & Uranus & J0303-6211 & 1.8 & 1.2 & 65 \\
2015-Dec-28 & 95.69 & 34 & 3.2$\times$3.5 & Uranus & J0309-6058 & 2.9 & 1.2 & 83 \\
2015-Dec-28 & 99.44 & 34 & 3.1$\times$3.4 & Uranus & J0309-6058 & 2.8 & 1.2 & 77 \\
2015-Dec-28 & 103.19 & 34 & 3.0$\times$3.4 & Uranus & J0309-6058 & 2.7 & 1.5 & 72 \\
2015-Dec-28 & 106.94 & 34 & 2.9$\times$3.3  & Uranus & J0309-6058 & 2.8 & 1.0 & 95 \\
\hline
\textbf{B6} & & & & & & & & \\
2016-Nov-03 & 233.65 & 45 & 0.25$\times$0.30 & J0334-4008 & J0303-6211 & 0.5 & 32.4 & 24 \\
\hline
\textbf{B7} & & & & & & & & \\
2016-Jun-04 & 343.48 & 41 &  0.31$\times$0.49 & J2258-2758 & J0303-6211 & 0.8 & 6.5 & 12\\
\hline
\textbf{B8} & & & 0.20$\times$0.30 & & & & & 53 \\
2016-Nov-15 & 423.63 & 41 &  & J0538-4405 & J0253-5441 & 0.8 & 11.4 & \\
2016-Nov-16 & 423.63 & 42 &  & J0538-4405 & J0253-5441 & 0.5 & 33.7 & \\
2016-Nov-16 & 423.63 & 42 &  & J0538-4405 & J0253-5441 & 0.4 & 33.7 & \\
2016-Nov-17 & 423.63 & 43 &  & J0538-4405 & J0253-5441 & 0.3 & 33.7 & \\
\hline
\end{tabular}

$^\mathrm a$~First local oscillator frequency\hspace{2em}
$^\mathrm b$~Precipitable water vapor at zenith\hspace{2em}
$^\mathrm c$~On-source integration time\hspace{2em} \\
$^\mathrm d$~Root-mean-square noise level in 7.5~GHz continuum image
\end{table}

\begin{table}
\centering
\caption{Optical/IR Photometry
\label{t:OIRdata}}
\scriptsize
\begin{tabular}{llccc}
\hline
Telescope & Instrument/Filter & Lens & {\sname}E & {\sname}W \\
\hline
\textit{HST} & ACS/F606W & $>$ 27.05 & $>$28.11 & $>$27.08 \\
\textit{HST} & ACS/F775W & $>$ 26.55 & $>$27.59 & $>$26.63 \\
Gemini & GMOS/$i'$ & 25.00$\pm$0.20 & & \\
Gemini & GMOS/$z'$ & 24.40$\pm$0.20 & & \\
\textit{HST} & WFC3/F125W & 23.06$\pm$0.16 & 25.28$\pm$0.10 & $>$26.69  \\
\textit{HST} & WFC3/F160W & 22.76$\pm$0.15 & 24.98$\pm$0.12 & $>$27.11 \\
Gemini & FLAMINGOS/$K_s$ 2.16~\um & 22.42$\pm$0.13 & ... & ...\\
\textit{Spitzer} & IRAC/Ch1 3.6~\um & 21.40$\pm$0.14 & 24.47$\pm$0.30 & (23.87$\pm$0.28) \\
\textit{Spitzer} & IRAC/Ch2 4.5~\um &21.63$\pm$0.13 &  24.45$\pm$0.25 & (23.63$\pm$0.22) \\
\hline
\end{tabular}\\
All data in apparent (i.e., not corrected for magnification) AB magnitudes. Limiting magnitudes reported as $1\sigma$ values. 
The magnification estimates for the E and W sources are 1.3 and 2.1, respectively, as reported in Methods Section~\ref{s:lensmodel}.
IRAC photometry for \sname{W} is uncertain due to blending with the lens, as noted in Methods Section~\ref{s:deblend}.
\end{table}

\begin{table}
\centering
\caption{Far-Infrared Photometry
\label{t:FIRdata}}
\scriptsize
\begin{tabular}{lcccc}
\hline
Telescope      & Observed Wavelength & $S_\nu$ (east intrinsic) & $S_\nu$ (west intrinsic) & $S_\nu$ (total apparent) \\
\hline
\textit{Herschel}/SPIRE$^\mathrm a$ & 250\,\um   &   1.9$\pm$0.6  & 12.7$\pm$4.2  & 29.0 $\pm$ 8.0  \\
\textit{Herschel}/SPIRE$^\mathrm a$ & 350\,\um   &   2.5$\pm$0.5  & 16.6$\pm$2.9  & 38.0 $\pm$ 6.0  \\
\textit{Herschel}/SPIRE$^\mathrm a$ & 500\,\um   &   3.5$\pm$0.6  &  22.7$\pm$4.2  & 52.0 $\pm$ 8.0  \\
ALMA/B8           & 710\,\um   & 3.1 $\pm$ 0.2     & 19.9 $\pm$ 0.3  &                 \\
ALMA/B7           & 869\,\um   & 2.9 $\pm$ 0.2     & 15.9 $\pm$ 0.25 &                 \\
ALMA/B6           & 1.26\,mm   & 1.18 $\pm$ 0.05   & 9.77 $\pm$ 0.15 &                 \\
ALMA/B3           & 3\,mm      & 0.040 $\pm$ 0.028 & 0.76 $\pm$ 0.02 &                 \\
\hline
\end{tabular}\\
Flux densities ($S_\nu$) in mJy. \\
$^\mathrm a$~\textit{Herschel} photometry does not spatially resolve the two components. See Methods Section~\ref{s:lensmodel} 
for details.
\end{table}
}
\renewcommand{\figurename}{Extended Data Figure}
\setcounter{figure}{0}

\begin{figure}
\includegraphics[trim=0in 0in 0in 0.0in, clip,width=6.5in]{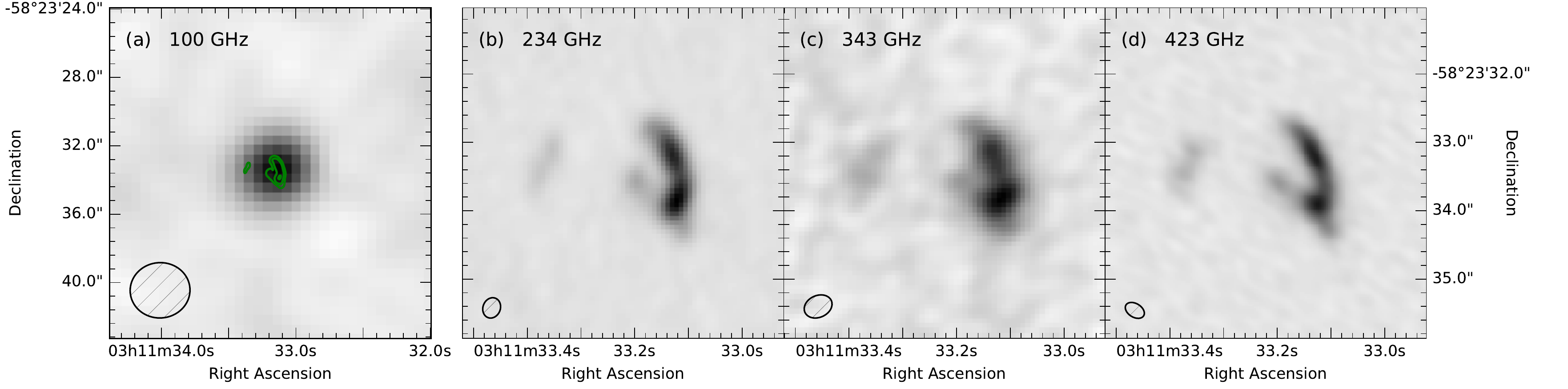} 
\caption{\textbf{ALMA Continuum Images of \sname.} 
\label{f:ALMAall}
From left to right, continuum images in ALMA bands 3, 6, 7, and 8, corresponding to rest-frame wavelengths of 380, 160, 110, and 90~\um, 
respectively. Note that the resolution in the first panel is a factor of roughly ten worse than the other images, and the displayed field of 
view is also larger by a factor of 4. Contours at 10, 30, and 90\% of the image peak in band 6 are shown in the band 3 panel for scale. 
The ALMA synthesized beam (full-width-at-half-maximum) is represented as a hatched ellipse in the corner of each image.}
\end{figure}

\begin{figure}
\includegraphics[trim=0.in 0.15in 0.0in 0.0in, clip,width=6.5in]{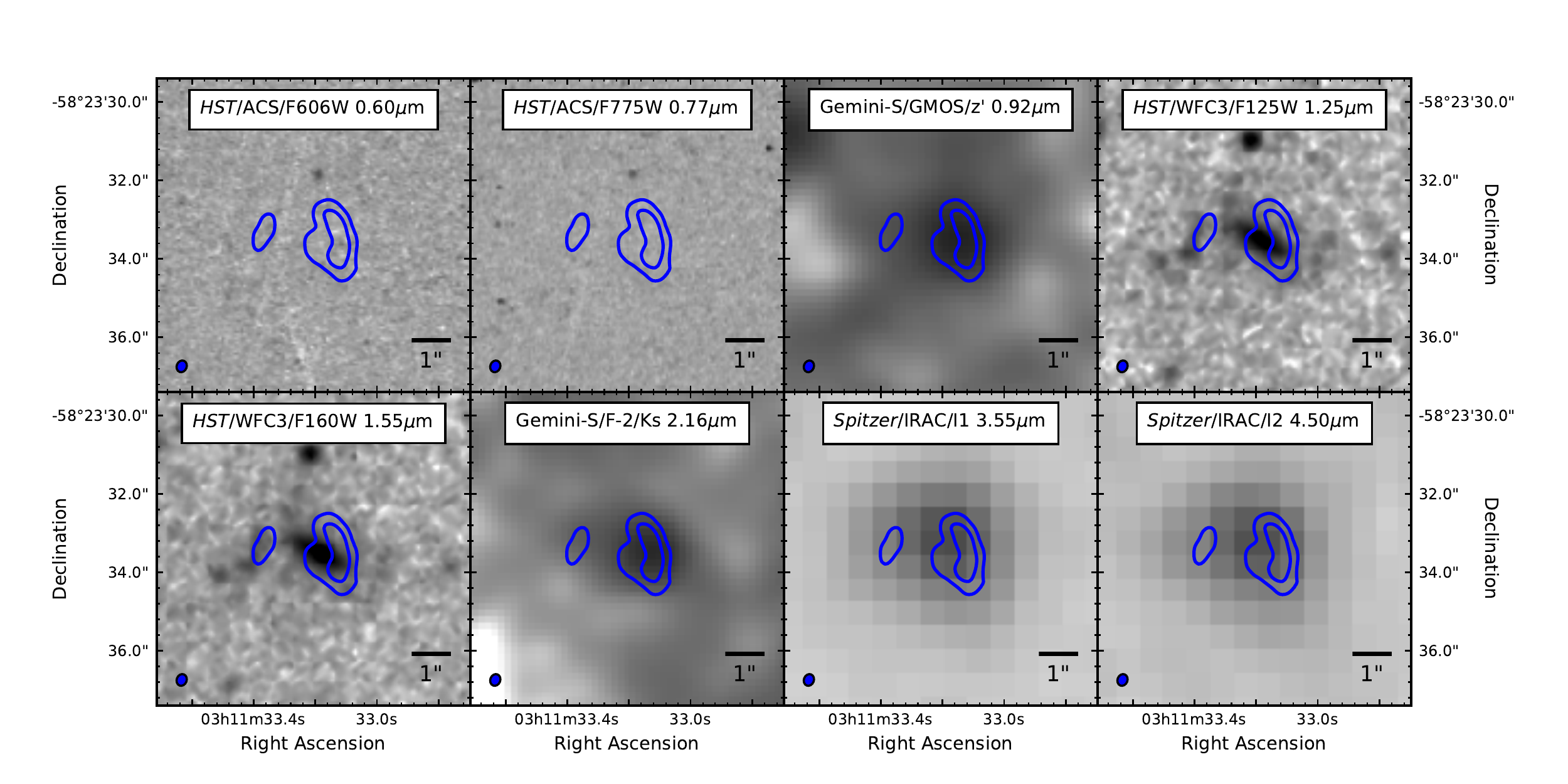} 
\caption{\textbf{Infrared and optical imaging of \sname.} 
\label{f:oir}
8\arcsec$\times$8\arcsec\ thumbnails of \sname\ in the observed optical and infrared filters.
ALMA band 6 continuum contours at 30 and 4\% of the image peak are shown in blue along with the 
ALMA synthesized beam depicted as blue ellipse in the corner of each image.
}
\end{figure}

\begin{figure}
\includegraphics[width=6.5in]{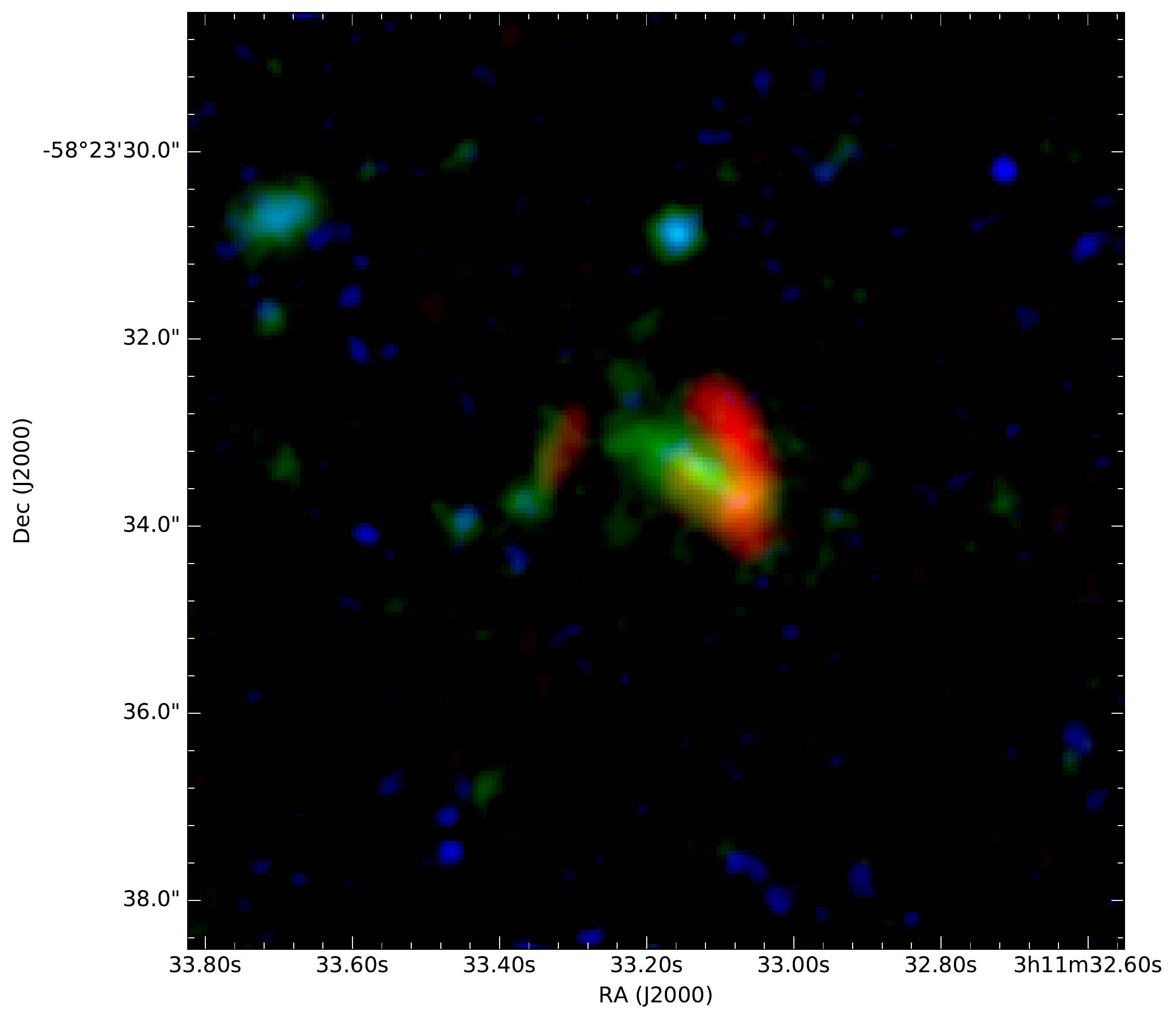} 
\caption{\textbf{Optical, Infrared, and Millimeter image of \sname.} 
\label{f:3color}
The field around \sname\ as seen with ALMA and \hst\ at 1.3~mm (ALMA band 6; red), 
1300~nm (combined \textit{Hubble}/WFC3 F125W and F160W filters; green), and
700~nm (combined \textit{Hubble}/ACS F606W and F775W filters; blue).
For emission from $z=6.9$, no emission should be visible in the ACS filters due to the opacity of the neutral intergalactic medium, 
while the other filters correspond to rest-frame 160~nm and 160~\um.
}
\end{figure}

\begin{figure}
\includegraphics[width=6.5in]{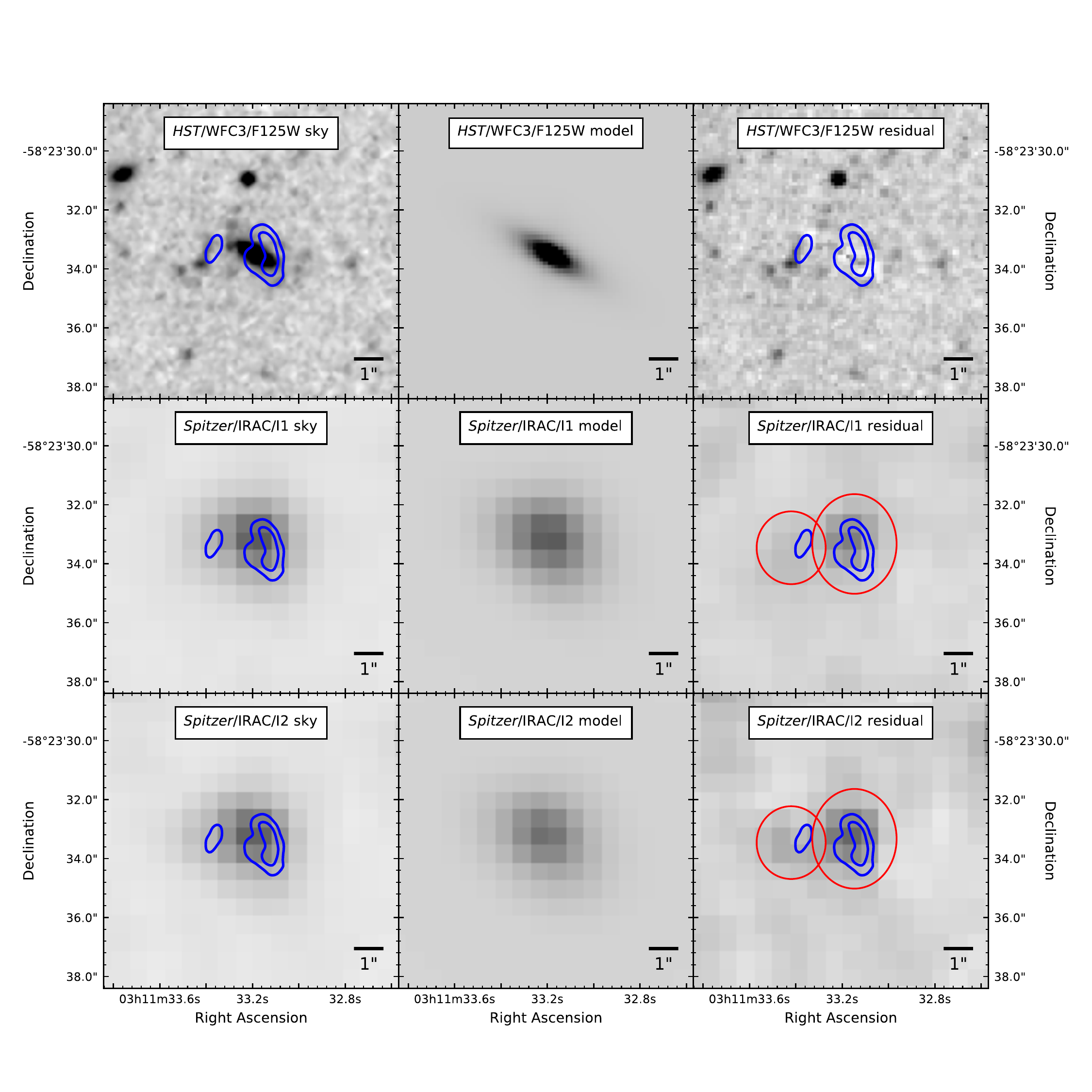} 
\caption{\textbf{De-blending of the optical/infrared images.} Left--right: sky image, model, and residual images. 
Top--bottom: \textit{Hubble}/WFC3 F125W, \textit{Spitzer}/IRAC 3.6~\um, and 4.5~\um\ data.
The ALMA band 6 contours are shown in the left and right columns, and the red circles in the right column show the 
photometric extraction regions for the \textit{Spitzer}/IRAC images.
\label{f:deblend}
}
\end{figure}


\begin{figure}
\centering
\includegraphics[width=\textwidth]{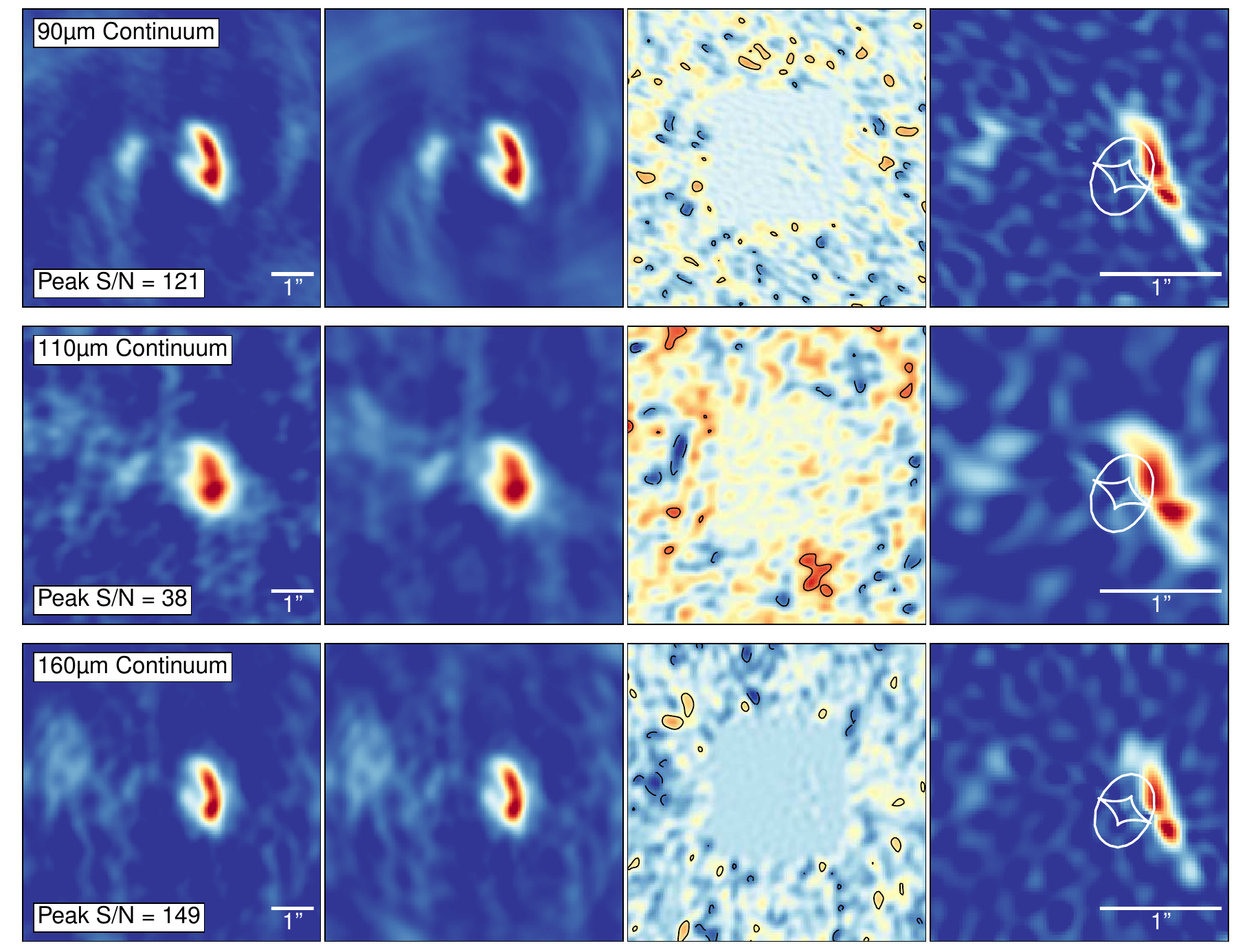}
\caption{\textbf{Gravitational lensing model of the dust continuum emission in \sname.}
\label{f:lenscont}
For each continuum wavelength for which we have suitable data, we reconstruct the source-plane emission as described in Section~\ref{s:lensmodel}. For each wavelength, from left to right, we show the ``dirty'' (i.e., not deconvolved) image of the data, the dirty image of the model, the model residuals, and the source-plane reconstruction. Because the images of the data are not deconvolved, the apparently high-significance structure far from the object is due to sidelobes in the synthesized beam, and should be reproduced by the models. The image-plane region modeled is evident in the residuals, and results in the ``noise'' in the source-plane reconstructions. Contours in the residual panels are drawn in steps of $\pm2\sigma$. The lensing caustics are shown in each source-plane panel. The lens parameters are determined independently at 90 and 160\,\um; at 110\,\um\ we adopt the parameters of the 160\,\um\ model.
}
\end{figure}

\begin{figure}
\centering
\includegraphics[width=0.9\textwidth]{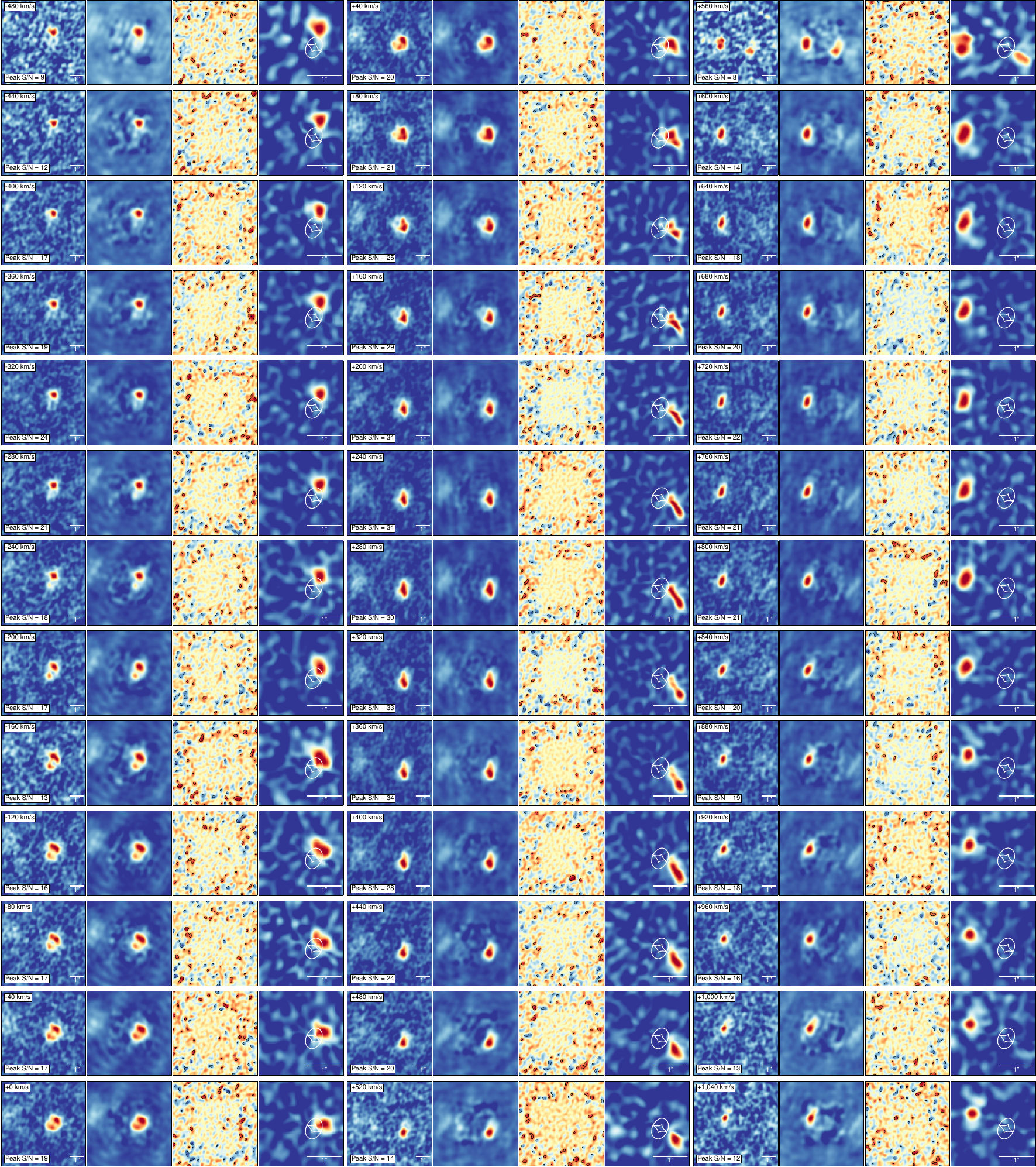}
\caption{\textbf{Gravitational lensing model of the \cii\ line in \sname.}
\label{f:lenscii}
For each 40~\kms-wide channel, we reconstruct the source-plane emission using the lens parameters determined from fitting to the rest-frame 160\,\um\ (ALMA Band 6) continuum data (Section~\ref{s:lensmodel}). We show for each channel, from left to right, the ``dirty'' (i.e., not deconvolved) image of the data, the dirty image of the model, the model residuals, and the source-plane reconstruction. Because the images of the data are not deconvolved, the apparently high-significance structure far from the object is due to sidelobes in the synthesized beam, and should be reproduced by the models. Contours in the residual panels are drawn in steps of $\pm2\sigma$. The lensing caustics are shown in each source-plane panel.
}
\end{figure}


\begin{figure}
\includegraphics[trim=0.in 0.in 0.3in 0.4in, clip,width=6.5in]{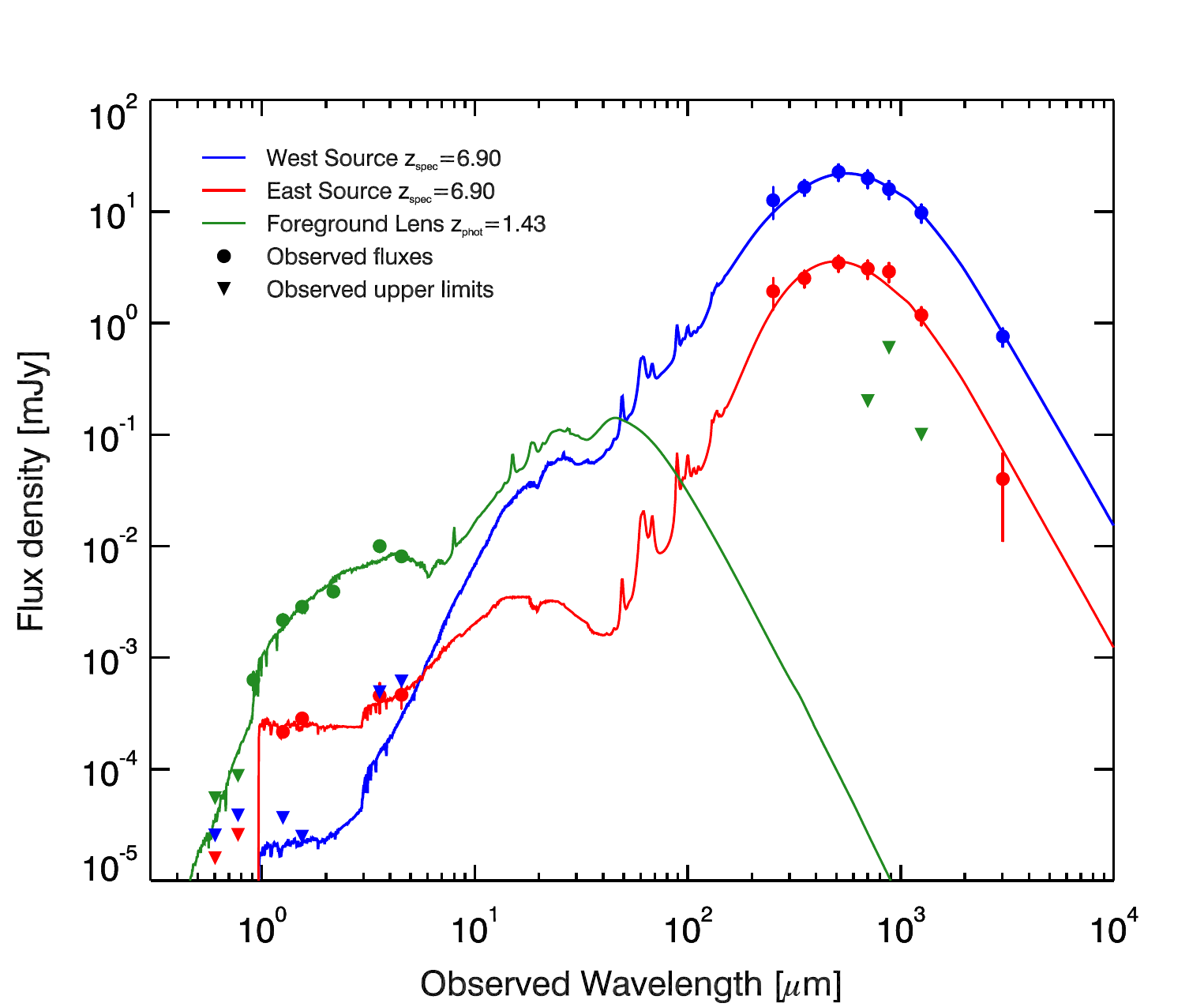}
\caption{\textbf{Optical to submillimeter spectral energy distribution modeling for \sname~E and W and the lensing galaxy.} 
\label{f:fullSED}
The photometric data in Extended Data Tables~\ref{t:OIRdata} and \ref{t:FIRdata} for the three components at the position of \sname\ 
are compared to the models determined using the CIGALE spectral energy distribution modeling code. The lens is modeled 
assuming a redshift of 1.43, as estimated with the photometric redshift code EAZY. Upper limits are shown at the 1$\sigma$ threshold.}
\end{figure}

\end{document}